\documentclass[twocolumn,pacs,floatfix,amsmath,nofootinbib,superscriptaddress,amssymb,preprintnumbers,floatfix]{revtex4}
\usepackage{amsmath,txfonts,longtable,lscape,booktabs,overpic,amssymb,bm,multirow,float,graphicx,color,dcolumn,hyperref}
\hypersetup{colorlinks,citecolor=blue,anchorcolor=red,menucolor=red, linkcolor=red,filecolor=red,runcolor=red,urlcolor=blue,frenchlinks=true}

\allowdisplaybreaks

\begin{document}

	\title{Schwinger pair production rate and time for some space-dependent electromagnetic fields via worldline instantons formalism}
	
	\author{Orkash Amat}
	\affiliation{ Key Laboratory of Beam Technology of the Ministry of Education, and College of Nuclear Science and Technology, Beijing Normal University, Beijing 100875, China}
	\affiliation{ Institute of Radiation Technology, Beijing Academy of Science and Technology, Beijing 100875, China
	}
	
	\author{Li-Na Hu}
	\affiliation{ Key Laboratory of Beam Technology of the Ministry of Education, and College of Nuclear Science and Technology, Beijing Normal University, Beijing 100875, China}
	\affiliation{ Institute of Radiation Technology, Beijing Academy of Science and Technology, Beijing 100875, China
	}
	
	\author{Adiljan Sawut}
	\affiliation{ School of Science, China University of Mining and Technology, Beijing 100083, China}

	\author{Melike Mohamedsedik}
	\affiliation{ Key Laboratory of Beam Technology of the Ministry of Education, and College of Nuclear Science and Technology, Beijing Normal University, Beijing 100875, China}
	\affiliation{ Institute of Radiation Technology, Beijing Academy of Science and Technology, Beijing 100875, China
	}
	
	\author{M. A. Bake}
    \affiliation{ School of Physics Science and Technology, Xinjiang University, Urumqi, Xinjiang 830046, China}

	\author{B. S. Xie \footnote{bsxie@bnu.edu.cn}}
	\affiliation{ Key Laboratory of Beam Technology of the Ministry of Education, and College of Nuclear Science and Technology, Beijing Normal University, Beijing 100875, China}
	\affiliation{ Institute of Radiation Technology, Beijing Academy of Science and Technology, Beijing 100875, China
	}
\begin{abstract}
Schwinger pair production in some of space-dependent electromagnetic fields is studied analytically by using worldline instantons formalism for scalar quantum electrodynamics. With the increase of the modified Keldysh parameter $\gamma_{keb}$, the pair production rate decreases, the region of instanton paths expands regardless of the electromagnetic field form. We also find that all of the paths of the instantons for various electromagnetic fields are exactly in the same plane while with the different elliptical shapes. Moreover, the magnitudes of the normalized instantons action with different external electromagnetic fields are all bounded within the region from $\pi$ to $2\pi$ in spatial inhomogeneity fields. We further analytically identify and obtain two kinds of the pair-production time associated to the worldline instantons action.
\end{abstract}
\pacs{12.20.Ds, 03.65.Pm, 02.60.-x}
\maketitle

\section{introduction}\label{sec:intro}
Vacuum pair production is not only one of the interesting non-perturbative phenomena in the quantum field theory (QFT), but also is a window into the strong field theory, and this phenomena has been studied with different methods. Dirac first put forward the relativistic wave equation and predicted the existence of positrons~\cite{Dirac:1928hu}. Klein calculated the transmission and reflection probabilities at a potential step using the Dirac equation~\cite{Klein:1929zz}. Sauter performed calculation for a piecewise linear potential which describes a constant and finite electric field over a finite length~\cite{Sauter:1931zz}. Heisenberg and Euler calculated the vacuum pair production rate ${\rm exp}(-\pi E_{cr}/E)$ for the electromagnetic fields and defined the critical field strength as $E_{cr}=m^{2}c^{3}/e\hbar \thickapprox 1.32 \times 10^{18}\rm{~V/m}$~\cite{Heisenberg:1936nmg}. It is a great challenge to achieve this extreme field experimentally for the present laser technology~\cite{Ringwald:2001ib,Heinzl:2009bmy,Marklund:2008gj}. Also, Schwinger formalized the vacuum pair creation process via quantum electrodynamics (QED) and recovered the pair production rate in a constant field with the proper-time approach~\cite{Schwinger:1951nm} as
\begin{align}
{\rm Im}\,{\cal L}[E]
&=\frac{e^2 E^2}{16\pi^3}\, \sum_{n=1}^\infty \frac{(-1)^{n-1}}{n^2}
\,\exp\left[-\frac{ n\pi m^2}{e E}\right].
\label{L1scalim}
\end{align}
This non-perturbative vacuum pair creation process is called Sauter-Schwinger effect or Schwinger pair production.

The concept of instantons was first proposed in Yang-Mills theory (see Ref.~\cite{Yang:1954ek}) and called Belavin-Polyakov-Schwartz-Tyupkin (BPST) instanton after Belavin, Polyakov, Schwartz and Tyupkin obtained a classical solution to the Euclidean field equations with finite action~\cite{Belavin:1975fg,Polyakov:1976fu}, but it is difficult to obtain the analytical solutions. Thus, a new analytical solution method, path integral approach, is developed by Feynman for non-relativistic quantum mechanics~\cite{Feynman:1950ir,Feynman:1951gn}. These solutions of the paths are called worldline instantons, because they are solved by path integral approach. The worldline instantons approach of QFT is a powerful analytical and computational method both for perturbative and non-perturbative phenomena in QED~\cite{tHooft:1976rip}.

In recent years, many studies on Sauter-Schwinger effect has been investigated by different methods. The current widely used methods include the real time Dirac-Heisenberg-Wigner (DHW) formalism~\cite{Hebenstreit:2011wk,Kohlfurst:2019mag,Aleksandrov:2019ddt,Ababekri:2019dkl,Olugh:2018seh,Li:2021wag,Li:2021vjf,Mohamedsedik:2021pzb,Kohlfurst:2021dfk}, the computational quantum field theory (CQFT)~\cite{Krekora:2004trv,PhysRevA.73.022114,Tang:2013qna,Wang:2019oyk,Wang:2021tmo}, the quantum Vlasov equation (QVE)~\cite{Kluger:1998bm,Schmidt:1998vi,Li:2014xga,Li:2014psw,Gong:2020jqs}, the Wentzel-Kramers-Brillouin (WKB) approach~\cite{Brezin:1970xf,Popov:1971iga,Kim:2007pm,Hebenstreit:2009km,Dumlu:2010ua,Strobel:2014tha,Oertel:2016vsg}, the worldline instanton technique~\cite{Dunne:2005sx,Dunne:2006st,Dunne:2006ur,Dunne:2006ff,Dunne:2008zza,Dunne:2008kc,Dumlu:2011cc,BaisongXie:2012,Ilderton:2015lsa,Schneider:2014mla,Xie:2017,Schneider:2018huk,Esposti:2021wsh,Rajeev:2021zae} and so on. However, most of these studies do not have analytical solutions and explanations, and most of the results of the vacuum pair productions caused by the external electromagnetic field are waiting for the analytical solutions and interpretations. At the same time, there are few information about the quantum tunneling time with the Schwinger effect in electromagnetic fields~\cite{Landauer:1994zz}. The worldline instanton approach may be the best candidate to investigate the quantum tunneling time. Because of this, we define the quantum tunneling time by using the one loop worldline instanton method via scalar quantum electrodynamics (sQED)~\cite{Dunne:2005sx}.

In this paper, we investigate how the results of pair production are  affected in the space-dependent inhomogeneous electromagnetic fields with the modified Keldysh parameter $\gamma_{keb}$ analytically. Furthermore, we discuss the paths, instanton action, effective action and tunneling time for various $\gamma_{keb}$. For the pair-production time, two kinds of definition, vacuum decay time and tunneling time, are obtained by the analytical expression associated to the worldline instanton formalism.

Our paper is organized as follows. In Sec.~\ref{sec:formula}, we briefly introduce the general form of the one loop worldline instanton formalism for sQED in the electromagnetic fields. In Sec.~\ref{model}, we consider the model for the space-dependent inhomogeneous electromagnetic fields, and discuss the instanton paths and instanton action for various electromagnetic fields. In Sec.~\ref{tunnelingtime} we analytically discuss the vacuum decay or/and tunneling time for the Schwinger effect. In Sec.~\ref{discussion}, we mainly study the variation of the instanton trajectories for different space-dependent inhomogeneous electromagnetic fields, and confirm all of the instanton paths in electromagnetic fields are in the same planar subspace in the $({\rm Im}x_{1},x_{3},x_{4})$ space. The summary and outlook are given briefly in Sec.~\ref{conclusion}.

\section{worldline instanton formalism}\label{sec:formula}

Our calculations start with effective action $\Gamma^{\rm Mink}$ in the Minkowski space, we define it as
\begin{align}\label{gamma}
e^{i \Gamma^{\rm Mink}}:&=\left\langle \mathcal{O_{\rm {out}}}~ \vert ~\mathcal{O_{\rm {in}}}  \right\rangle,
\end{align}
 where the subscripts in and out represent the initial and final states of vacuum, $\left\langle \mathcal{O_{\rm {out}}}~ \vert ~\mathcal{O_{\rm {in}}}  \right\rangle$ is the vacuum persistence amplitude, and the pair production probability can be written as
\begin{align}\label{probability}
\mathcal{P}&=1 - \vert \left\langle \mathcal{O_{\rm {out}}}~ \vert ~\mathcal{O_{\rm {in}}}  \right\rangle \vert ^2=1-e^{-2 {\rm Im} \Gamma^{\rm Mink}}\approx 2 {\rm Im} \Gamma^{\rm Mink},
\end{align}
where ${\rm Im} \Gamma^{\rm Mink} \ll 1$. The relationship between Minkowski effective action and Euclidean effective action $\Gamma^{\rm Mink}=i \Gamma^{\rm Eucl}$ is given in Ref.~\cite{Dunne:2006st}, so that we can obtain
\begin{align}\label{im-re}
{\rm Im} \Gamma^{\rm Mink}&={\rm Re} \Gamma^{\rm Eucl},
\end{align}
i.e., we are going to Wick-rotate ${\rm Im} \Gamma^{\rm Mink}$ from Minkowski space to Euclidean space in order to simplify the path integral via $x_{4}=i t$, and we use the ${\rm Re} \Gamma^{\rm Eucl}$ to compute the pair production probability. Now the Minkowski four-potential $A^{\rm M}_{\mu}=\left(\phi,~{\bm A}^{\rm M}\right)=\left(A^{\rm M}_{0},~A^{\rm M}_{1},~A^{\rm M}_{2},~A^{\rm M}_{3}\right)$ turns into the Euclidean four-potential $A^{\rm E}_{\mu}=\left(A^{\rm E}_{4},~A^{\rm E}_{1},~A^{\rm E}_{2},~A^{\rm E}_{3}\right)=\left(A_{4},~{\bm A}\right)$. The relationship between each component in the Minkowski space and the Euclidean space  can be written as
\begin{align}\label{mikoeucl}
A^{\rm E}_{4}&=\frac{1}{i}A_{0}\left(t=-ix_{4}\right),\\
A ^{\rm E}_{j}&={\bm A}\left(t=-ix_{4}\right),
\end{align}
where index $j=1,2,3$ mean the $x$, $y$ and $z$ direction in the position space, and we denote that $x=x_{1}$, $y=x_{2}$ and $z=x_{3}$.

The one loop Euclidean effective action in an Abelian spacetime dependent background gauge field $A^{E}_{\mu}$ for the scalar particle is written by the worldline path integral form \cite{Dunne:2005sx}
\begin{align}\label{effectiveaction}
\Gamma^{E}[A]& \simeq \sqrt{\frac{2\pi}{m} } \int \mathcal{D} x \frac{1}{\left(\int_{0}^{1}du \dot{x}^{2} \right)^{\frac{1}{4}}} e^{-\left(m \sqrt{\int_{0}^{1}du \dot{x}^{2}}+ie \int_{0}^{1}du {\bm A} \cdot  \dot{\bm x} \right)},
\end{align}
where the functional integral $\int \mathcal{D} x$ contains all closed spacetime paths $x^{\mu}(u)$ with period 1. The Eq.~\eqref{effectiveaction} satisfies the weak-field condition as
\begin{align}\label{weakfieldcondition}
m \sqrt{\int_{0}^{1}du \dot{x}^{2}} \gg 1.
\end{align}
The worldline action (instanton action~\cite{Dunne:2005sx}) $\mathcal{S}_{0}$ which is nonlocal defined as
\begin{align}\label{worldlineaction}
\mathcal{S}_{0}&= m \sqrt{\int_{0}^{1}du \dot{x}^{2}}+ie \int_{0}^{1}du {\bm A} \cdot  \dot{\bm x}.
\end{align}
The path $x_{\mu}(u)$ satisfies nonlinear differential equations system~\cite{Dunne:2005sx}
\begin{align}\label{eom}
m \frac{\ddot{x}_{\mu} }{\sqrt{\int_{0}^{1}du \dot{x}^{2}}}&=ieF_{\mu\nu}\dot{x}_{\nu},
\end{align}
where $m$ is instanton mass, $-e$ is the electron charge, the field strength $F_{\mu\nu}=\partial_{\mu}A_{\nu}-\partial_{\nu}A_{\mu}$ is an antisymmetric tensor, . Solutions of Eq.~\eqref{eom} which satisfies the periodicity condition $x_{\mu}(0)=x_{\mu}(1)$ are called worldline instantons. We can get
\begin{align}\label{constent}
\dot{x}^{2}={\rm constant} \equiv a^{2},
\end{align}
and the stationary instanton path $x_{\mu}(u)$ satisfies Eq.~\eqref{constent} for any $F_{\mu\nu}(x)$. Thus, the Eq.~\eqref{weakfieldcondition} can be rewritten as $ma \gg 1$. Therefore, we just deal with worldline action $\mathcal{S}_{0}$ to calculate one loop Euclidean effective action $\Gamma^{\rm E}[A]$. The Eqs.~\eqref{effectiveaction}, \eqref{worldlineaction} and \eqref{eom} can be written as
\begin{align}\label{eom2}
&m \ddot{x}_{\mu}=ieaF_{\mu\nu}\dot{x}_{\nu},\\ \label{eom22}
&\mathcal{S}_{0}= m a+ie \int_{0}^{1}du {\bm A} \cdot  \dot{\bm x},\\ \label{eom222}
&\Gamma^{E}[A] \simeq \sqrt{\frac{2\pi}{ma} } \int \mathcal{D} x  e^{-\mathcal{S}_{0}}.
\end{align}
It is not only easy to deal with above three equations, but also to get the basic physical informations of the pair production we need. In the next section, we discuss how to deal with space-dependent electromagnetic fields.

\section{spatially inhomogeneous electromagnetic fields}\label{model}

In this section, we solve instanton action and effective action analytically in a series of classical space-dependent background electromagnetic fields.

We choose a four-potential $A_{\mu}$ which has nonzero two components, and it is functions of $x_{3}$ as
\begin{align}\label{potential}
\begin{split}
A_{\mu}&=A^{\rm E}_{\mu}=\left(A^{\rm E}_{4},~A^{\rm E}_{1},~0,~0\right)\\ &=\left(-iA^{\rm M}_{0}(x_{3}),~A^{\rm M}_{1}(x_{3}),~0,~0\right).
\end{split}
\end{align}
 From Eq.~\eqref{potential}, the electromagnetic fields can be written as
\begin{align}\label{maxwellequatione}
{\bm E}&=-\bigtriangledown \phi - \frac{\partial {\bm A}}{\partial t}=E_{z}(x_{3}) {\bm e}_{z},\\ \label{maxwellequationb}
{\bm B}&=\bigtriangledown \times {\bm A}=B_{y}(x_{3}) {\bm e}_{y},
\end{align}
where ${\bm e}_y$ and ${\bm e}_z$ are the unit vectors, $E_{z}(x_{3})$ and $B_{y}(x_{3})$ are the magnitudes of the electric fields in the $z$ direction and the magnetic fields in the $y$ direction. In our cases, we compute paths $x_{\mu}$, worldline action $\mathcal{S}_{0}$, and discuss the effective action $\Gamma^{E}[A]$ for the four-potential $A_{\mu}$. Finally, we can get the equations of motion
\begin{align}\label{eom3}
m \ddot{x}_{1}&=-iae \frac{\partial A_{1}(x_{3})}{\partial x_{3}}\frac{\partial x_{3}}{\partial u},\\
m \ddot{x}_{2}&=ieaF_{2\nu}\dot{x}_{\nu}=0,\\
m \ddot{x}_{3}&=iae \left(\frac{\partial A_{1}(x_{3})}{\partial x_{3}}\frac{\partial x_{1}}{\partial u}+\frac{\partial A_{4}(x_{3})}{\partial x_{3}}\frac{\partial x_{4}}{\partial u}\right),\\
m \ddot{x}_{4}&=-iae \frac{\partial A_{4}(x_{3})}{\partial x_{3}}\frac{\partial x_{3}}{\partial u}.
\end{align}
If we assume that $A_{1}(x_{3})$ and $A_{4}(x_{3})$ are odd functions of the $x_{3}$, we get $\dot{x}^{2}_{2}=0$ and $\dot{x}^{2}_{1}+\dot{x}^{2}_{3}+\dot{x}^{2}_{4}=a^{2}$ from above equations, which leads to the result
\begin{align}\label{differentialequations1}
\dot{x}_{1}&=- \frac{iae}{m} A_{1}(x_{3}),\\ \label{differentialequations2}
\dot{x}_{4}&=- \frac{iae}{m} A_{4}(x_{3}),\\ \label{differentialequations3}
|\dot{x}_{3}|&=a \sqrt{1+\left(\frac{e}{m}A_{1}(x_{3})\right)^{2}+\left(\frac{e}{m}A_{4}(x_{3})\right)^{2}}.
\end{align}
When $\dot{x}_{3}=0$ in Eq.~\eqref{differentialequations3}, we can get
\begin{align}\label{turningpoints}
1+\left(\frac{e}{m}A_{1}(x^{*}_{3})\right)^{2}+\left(\frac{e}{m}A_{4}(x^{*}_{3})\right)^{2}\equiv 0.
\end{align}
At the turning point $x^{*}_{3}$, the kinetic energy equals to potential energy of the instanton (see Ref.~\cite{Schneider:2019}). It is vital to calculate the path integral in the effective action.

Note that the resulting worldline action is obtained by taking Eq.~\eqref{potential} into Eq.~\eqref{eom22} as
\begin{align}\label{worldline0}
\begin{split}
\mathcal{S}_{0}&=ma + ie \int_{0}^{1}du \left(A_{1}   \dot{x}_{1} +A_{4}   \dot{x}_{4}\right)\\
&=ma + ie \int_{0}^{1}du \left(\frac{im}{ae}  \dot{x}^{2}_{1} + \frac{im}{ae} \dot{x}^{2}_{4}\right)\\
&=ma+\frac{m}{a}\int_{0}^{1}du \left({x}^{2}_{1}+\dot{x}^{2}_{4}\right)\\
&=ma-\frac{m}{a}\int_{0}^{1}du \left(a^{2}-\dot{x}^{2}_{3}\right)\\
&=\frac{m}{a}\int_{0}^{1}du \dot{x}^{2}_{3}.
\end{split}
\end{align}
In general, the expressions of $A_{1}(x_{3})$ and $A_{4}(x_{3})$ in the Euclidean space can be written as
\begin{align}\label{twopotentials}
A_{1}(x_{3})&=\frac{B}{k} f(kx_{3}),\\ \label{twopotentials11}
A_{4}(x_{3})&=-\frac{iE}{k} f(kx_{3}),
\end{align}
where $i$ is the imaginary unit, $k$ is the field frequency with space, $1/k$ characterize the length scale of the spatial inhomogeneity, $E$ and $B$ are peak values of the electric and magnetic fields strength, respectively. Note that $f$ is the odd function with respect to $x_{3}$.

After substituting Eqs.~\eqref{twopotentials} and~\eqref{twopotentials11} into the differential equations~\eqref{differentialequations1}, ~\eqref{differentialequations2} and ~\eqref{differentialequations3}, we get
\begin{align}\label{differentialequations11}
\dot{x}_{1}&=- \frac{iaeB}{mk} f(kx_{3}),\\ \label{differentialequations21}
\dot{x}_{4}&=- \frac{aeE}{mk} f(kx_{3}),\\ \label{differentialequations31}
|\dot{x}_{3}|&=a \sqrt{1- \frac{f^{2} (kx_{3})}{\gamma^{2}_{keb}}}.
\end{align}
For a convenience, we define a modified Keldysh parameter that $\gamma_{keb}=mk/e\sqrt{E^2-B^2}$, which implies that $\gamma^{-2}_{keb}=\left(\gamma^{-2}_{ke}-\gamma^{-2}_{kb}\right)$, where $\gamma_{ke}=mk/eE$ and $\gamma_{kb}=mk/eB$. The instanton action can be written as
\begin{align}\label{worldline01}
\begin{split}
\mathcal{S}_{0}&=\frac{m}{a}\int_{0}^{1}du \dot{x}^{2}_{3}=ma\int_{0}^{1}du \left(1- \frac{f^{2} (kx_{3})}{\gamma^{2}_{keb}}\right)\\
&=\frac{4mn}{k\gamma_{keb}} \int_{0}^{1}dy \frac{\sqrt{1-y^{2}}}{|f^{'}|}\equiv \frac{\pi mn}{k\gamma_{keb}} g(\gamma_{keb}),
\end{split}
\end{align}
where $y=\frac{1}{\gamma_{keb}} f(\upsilon)$ with $\upsilon= k x_{3}=f^{-1}(\gamma_{keb} y)$, thus, $f^{'}(\upsilon)$ can be reexpressed as a function of $y$ again, and the function $g(\gamma_{keb})$ is defined as
\begin{align}\label{g}
g(\gamma_{keb})&\equiv \frac{2}{\pi} \int_{-1}^{1}dy \frac{\sqrt{1-y^{2}}}{|f^{'}|},
\end{align}
where integral bounds are related to the turning points \cite{Schneider:2019}
\begin{align}\label{TPs}
x_{3}^{*}=\pm \frac{f^{-1}(\gamma_{keb})}{k}.
\end{align}
In the next subsections, we compare these instanton action in Eqs.~\eqref{worldline0} and \eqref{worldline01} in order to verify whether the two results are equivalent.
\subsection{Constant field}
First, we begin with investigating the instanton action for constant electromagnetic background fields. If $f(kx_{3})=kx_{3}$ and $k=1$, then we can obtain the Euclidean four-potential for constant electromagnetic background fields
\begin{align}\label{Amueb}
A^{E}_{\mu}&=\left(-iEx_{3},Bx_{3},0,0\right).
\end{align}
The solutions of the Eqs.~\eqref{differentialequations11}, \eqref{differentialequations21} and~\eqref{differentialequations31} are
\begin{align}\label{patheb1}
x_{1}&=\frac{i\gamma^{2}_{keb}}{\gamma_{kb}} {\rm cos}(2n\pi u),\\ \label{patheb2}
x_{3}&=\gamma_{keb} {\rm sin}(2n\pi u),\\ \label{patheb3}
x_{4}&=\frac{\gamma^{2}_{keb}}{\gamma_{ke}} {\rm cos}(2n\pi u),
\end{align}
where
\begin{align}\label{aeb}
a&=2n\pi \gamma_{keb}, ~n\in Z^{+},
\end{align}
$n$ is integer number of the closed paths.

\begin{figure}[ht!]\centering
\includegraphics[width=0.45\textwidth]{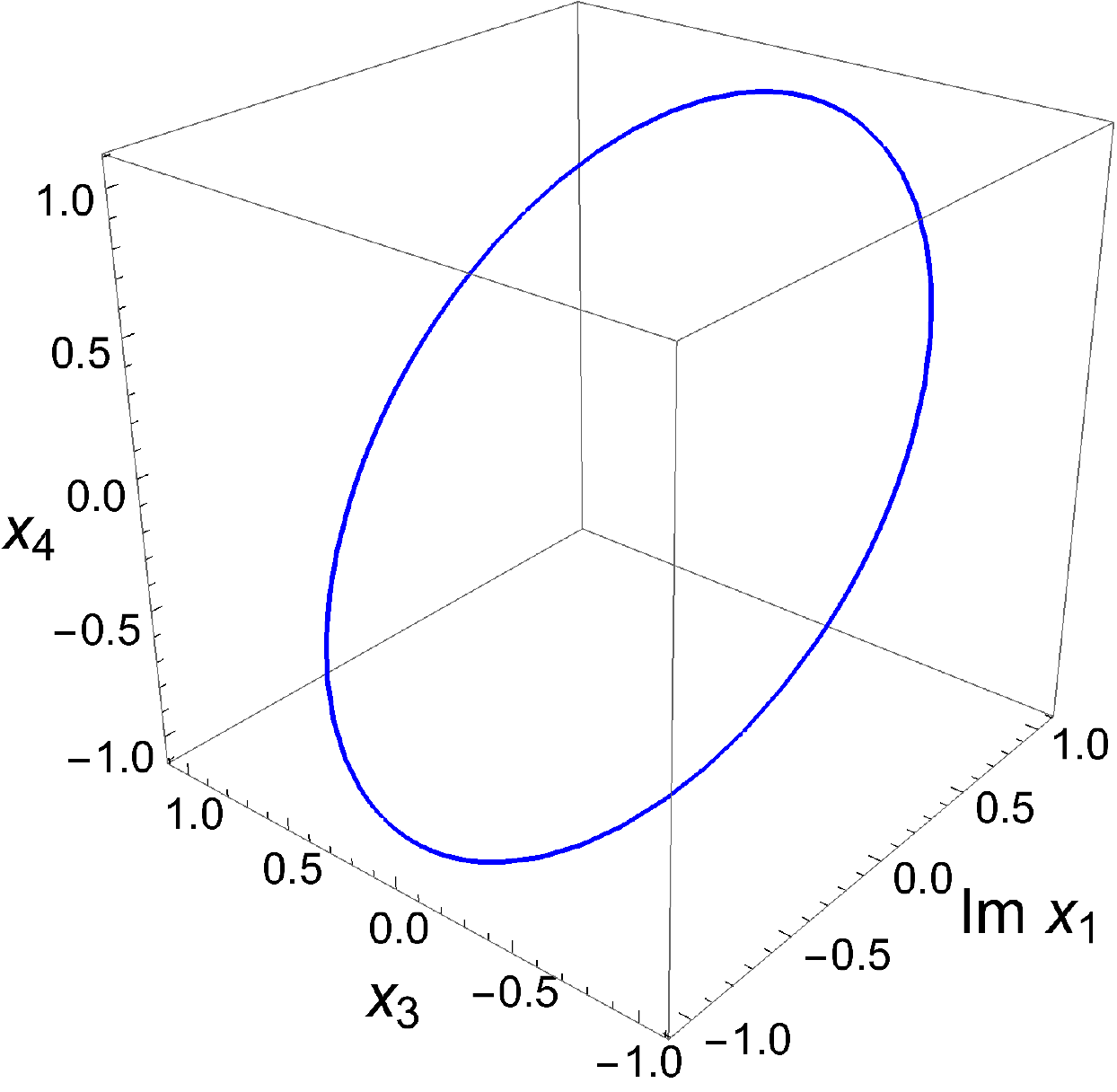}
\caption{Parametric plot of the stationary worldline instanton paths in the $({\rm Im}x_{1},~x_{3},~x_{4})$ space for the cases of constant electric and magnetic fields of strength $E$ and $B$. The paths are elliptic curves in $({\rm Im} x_{1},~x_{3},~x_{4})$ space in units of $mB/e(E^{2}-B^{2})$, $m/e\sqrt{E^{2}-B^{2}}$ and $mE/e(E^{2}-B^{2})$ for ${\rm Im} x_{1}$, $x_{3}$ and $x_{4}$.
\label{fig:constantpaths}}
\end{figure}

The stationary worldline instanton paths are the elliptic curves with different $\gamma_{keb}$, as shown in Fig.~\ref{fig:constantpaths}. We find the same trajectories for different $\gamma_{keb}$ in $({\rm Im}x_{1},x_{3},x_{3})$ space. For $B=0$, we can obtain the same result with Eq.~(26) in Ref.~\cite{Dunne:2005sx}.

By taking the Eq.~\eqref{aeb} into the Eq.~\eqref{weakfieldcondition}, we get the weak field condition
\begin{align}\label{1}
\frac{2n\pi m^{2}}{e\sqrt{E^{2}-B^{2}}} \gg 1.
\end{align}
The physical meaning of the Eq.~\eqref{1} is that $\sqrt{E^{2}-B^{2}} \ll \frac{2\pi m^{2}c^{3}}{e \hbar }\sim 10^{16}{\rm V/cm}=E_{cr}$ or $\sqrt{E^{2}-B^{2}} \ll \frac{2\pi m^{2}c^{2}}{e \hbar }\sim 10^{9}{\rm T}=B_{cr}$, which is satisfied for experimentally accessible electromagnetic fields, and the result is the same with Eq.(25) in Refs.~\cite{Schwinger:1951nm,Dunne:2005sx} when $B=0$.

The corresponding instanton action by using Eq. \eqref{worldline0} is
\begin{align}\label{S0eb}
\mathcal{S}_{0}&=2n\pi m \gamma_{keb}\int_{0}^{1}du ~{\rm cos}^{2}(2n\pi u)=\frac{n\pi m^{2}}{e\sqrt{E^{2}-B^{2}}}.
\end{align}
Obviously, the instanton action $\mathcal{S}_{0}$ obtained from Eqs.~\eqref{worldline01} and \eqref{S0eb} have the same results with that in Refs.~\cite{Schwinger:1951nm,Dunne:2005sx}, which guarantee the correctness of our definition of the instanton action in Eq.~\eqref{worldline01}. The pair production does not occur when $B \geq E$ or $\gamma_{keb}$ is imaginary number. Therefore, we only consider the pair creation cases when $\gamma_{keb}$ is real number.
\subsection{Hyperbolic secant field}\label{sechfield}
For the Minkowski space-dependent electric field $E_{z}(x_{3})=E{\rm sech}^{2}(kx_{3})$ and  magnetic field $B_{y}(x_{3})=B{\rm sech}^{2}(kx_{3})$, the corresponding Euclidean space-dependent gauge potential is
\begin{align}\label{secheb}
A_{\mu}(x_{3})&=\left(-i\frac{E}{k}{\rm tanh}(kx_{3}),~\frac{B}{k}{\rm tanh}(kx_{3}),~0,~0\right).
\end{align}
By using the Eqs.~\eqref{differentialequations1}, \eqref{differentialequations2} and \eqref{differentialequations3}, we can get the following equations
\begin{align}\label{sech1}
\dot{x}_{1}&=- \frac{iae}{m} \frac{B}{k}\tanh(kx_{3}),\\ \label{sech2}
\dot{x}_{4}&= - \frac{ae}{m} \frac{E}{k}\tanh(kx_{3}),\\ \label{sech3}
|\dot{x}_{3}|&=a \sqrt{1+\left(\frac{e}{m} \frac{B}{k}\tanh(kx_{3})\right)^{2}-\left(\frac{e}{m}\frac{E}{k}\tanh(kx_{3})\right)^{2}}.
\end{align}
And the stationary solution for $x_{3}(u)$ is determined by integrating the Eq.~\eqref{sech3}
\begin{align}
x_{3}&=\frac{1}{k} {\rm arcsinh} \left(\frac{\gamma_{keb}}{\sqrt{1-\gamma^{2}_{keb}}} \sin\left( \frac{\sqrt{1-\gamma^{2}_{keb}}}{\gamma_{keb}} kau \right) \right).
\end{align}
We know that $\dot{x}^{2}_{1}+\dot{x}^{2}_{3}+\dot{x}^{2}_{4}=a^{2}$ is satisfied. Therefor, the $a$ can be obtained as
\begin{align}\label{secha}
a=\frac{\gamma_{keb}}{k\sqrt{1-\gamma^{2}_{keb}}} 2\pi n, ~n\in Z^{+}.
\end{align}
For this inhomogeneous case, we can rewrite Eq.~\eqref{weakfieldcondition} as
\begin{align}
\frac{m\gamma_{keb}}{k\sqrt{1-\gamma^{2}_{keb}}} 2\pi n \gg 1.
\end{align}
It is a weak-field condition, and  must satisfies $E>B$ and $\sqrt{E^{2}-B^{2}} \ll E_{cr}$ or $B_{cr}$.

\begin{figure}[ht!]\centering
\includegraphics[width=0.45\textwidth]{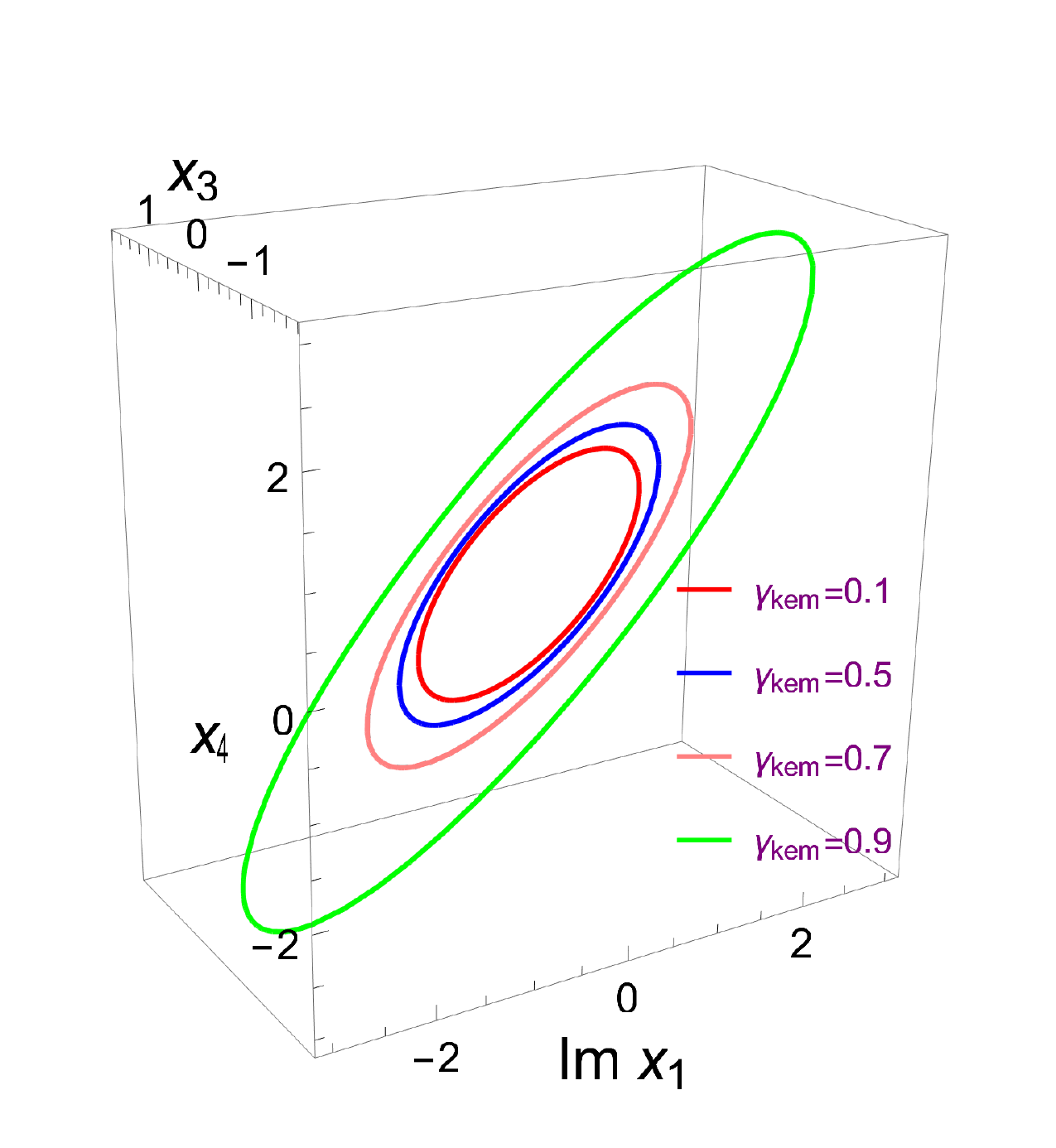}
\caption{Parametric plot of the stationary worldline instanton paths in the $({\rm Im}x_{1},~x_{3},~x_{4})$ space for the cases of the electric fields $E(x_{3})=E {\rm sech}^{2}(kx_{3})$ and the magnetic fields $B(x_{3})=B {\rm sech}^{2}(kx_{3})$ of strength $E$ and $B$. The paths are shown for various values of the parameter $\gamma_{keb}$ (red, blue, pink and green for $0.1$, $0.5$, $0.7$ and $0.9$), ${\rm Im} x_{1}$, $x_{3}$ and $x_{4}$ have been expresses in units of $mB/e(E^{2}-B^{2})$, $m/e\sqrt{E^{2}-B^{2}}$ and $mE/e(E^{2}-B^{2})$.
\label{fig:sechpath}}
\end{figure}

The periodic stationary instanton paths can be obtained from Eqs.~\eqref{sech1}, \eqref{sech2} and \eqref{sech3}
\begin{align}\label{pathsech1}
x_{1}&=\frac{i}{k}\frac{\gamma_{keb}}{\gamma_{kb}\sqrt{1-\gamma^{2}_{keb}}} {\rm arcsin}\left[ \gamma_{keb} ~\cos \left( 2\pi nu\right) \right],\\ \label{pathsech2}
x_{3}&=\frac{1}{k} {\rm arcsinh}\left[ \frac{\gamma_{keb}}{\sqrt{1-\gamma^{2}_{keb}}}  ~\sin \left( 2\pi nu\right) \right],\\ \label{pathsech3}
x_{4}&=\frac{1}{k}\frac{\gamma_{keb}}{\gamma_{ke}\sqrt{1-\gamma^{2}_{keb}}} {\rm arcsin}\left[ \gamma_{keb} ~\cos \left( 2\pi nu\right) \right].
\end{align}
The stationary worldline instanton paths are elliptic curves for the different $\gamma_{keb}$ as shown in Fig.~\ref{fig:sechpath}. As $\gamma_{keb}$ increases, the instanton paths shrinks in size.
Note that, when $\gamma_{keb} \rightarrow 0$, these loop trajectories of the instantons tend to become the same elliptic curves as shown in Fig. \ref{fig:constantpaths}. However, the loop paths expand enormously for $\gamma_{keb}=1$, and no pair creation in this case. Because the spatial width of the electromagnetic field is smaller than the Compton wavelength (see Ref.\cite{Dunne:2005sx}), thus, there are pair creation only for $\gamma_{keb}<1$. We can find that all of the elliptical shapes are in the same plane for any value of $\gamma_{keb}$. The stationary instanton paths have the same result with Eq.(61) in Ref.~\cite{Dunne:2005sx} when $B=0$.

\begin{figure}[ht!]\centering
\includegraphics[width=0.45\textwidth]{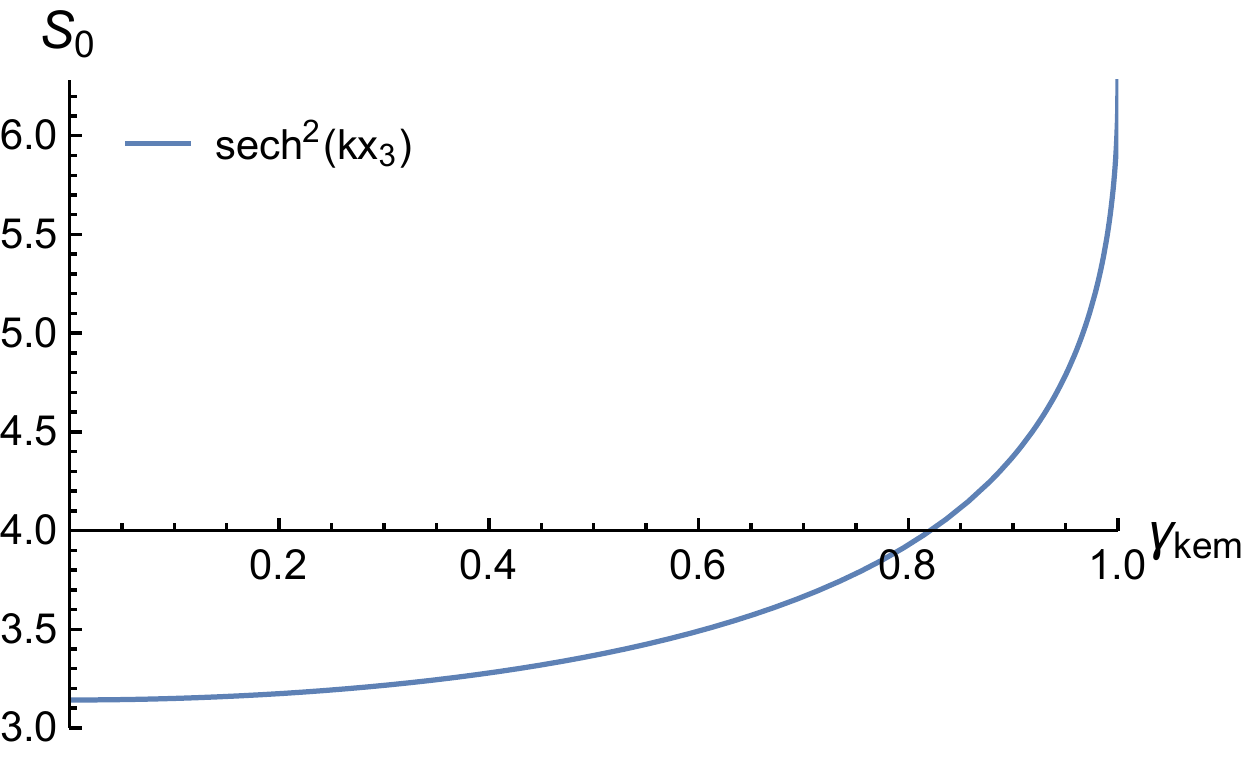}
\caption{Plot of the stationary worldline instanton action $S_{0}$ in the $({\rm Im}x_{1},~x_{3},~x_{3})$ space for the cases of the electric fields $E(x_{3})=E {\rm sech}^{2}(kx_{3})$ and the magnetic fields $B(x_{3})=B {\rm sech}^{2}(kx_{3})$ of strength $E$ and $B$. The $S_{0}$ have been expresses in units of $nm^{2}/e\sqrt{(E^{2}-B^{2})}$.
\label{fig:sechaction}}
\end{figure}

The stationary instanton action $\mathcal{S}_{0}$ is
\begin{align}\label{actionsech}
\begin{split}
\mathcal{S}_{0}&=ma \int_{0}^{1}du \frac{{\rm cos}^{2}\left( 2\pi nu \right)}{1+\frac{\gamma^{2}_{keb}}{1-\gamma^{2}_{keb}}} {\rm sin}^{2}\left( 2\pi nu \right)\\
&=\frac{\pi nm\gamma_{keb}}{k} \left( \frac{2}{1+\sqrt{1-\gamma^{2}_{keb}}} \right)\\
&=\frac{\pi nm^{2}}{e \sqrt{E^{2}-B^{2}}} \left( \frac{2}{1+\sqrt{1-\gamma^{2}_{keb}}} \right).
\end{split}
\end{align}
This stationary instanton action is plotted in Fig.~\ref{fig:sechaction}. Note that the instantons action $\mathcal{S}_{0}$ increases with $\gamma_{keb}$, and $S_{0}=\pi$ and $2\pi$ for $\gamma_{keb}=0$ and $1$, respectively. We can also obtain the same result for instanton action $\mathcal{S}_{0}$ with in Ref.~\cite{Dunne:2005sx}  when $B=0$.
\subsection{Cosine field}\label{cosfield}
In the Minkowski space we consider the electric field $E_{z}(x_{3})=E\cos(kx_{3})$  and  magnetic field $B_{y}(x_{3})=B\cos(kx_{3})$, and  corresponding Euclidean space gauge potential is
\begin{align}\label{coseb}
A_{\mu}(x_{3})&=\left(-i\frac{E}{k}\sin(kx_{3}),~\frac{B}{k}\sin(kx_{3}),~0,~0\right).
\end{align}
By using Eq.~\eqref{coseb}, we can obtain the following equations from Eqs.~\eqref{differentialequations1}, ~\eqref{differentialequations2} and ~\eqref{differentialequations3}
\begin{align}\label{cos1}
\dot{x}_{1}&=- \frac{iae}{m} \frac{B}{k}\sin(kx_{3}),\\ \label{cos2}
\dot{x}_{4}&= - \frac{ae}{m} \frac{E}{k}\sin(kx_{3}),\\ \label{cos3}
|\dot{x}_{3}|&=a \sqrt{1+\left(\frac{e}{m} \frac{B}{k}\sin(kx_{3})\right)^{2}-\left(\frac{e}{m}\frac{E}{k}\sin(kx_{3})\right)^{2}}.
\end{align}
The stationary $x_{3}(u)$ is determined by integrating Eq.~\eqref{cos3}
\begin{align}
du&=\frac{1}{a}\frac{dx_{3}}{\sqrt{1-\frac{{\rm sin}^{2}(kx_{3})}{\gamma^{2}_{keb}}}},
\end{align}
which has the solution
\begin{align}
x_{3}&=\frac{1}{k} {\rm arcsin} \left[\frac{\gamma_{keb}}{\sqrt{1-\gamma^{2}_{keb}}} {\rm sd}\left( \frac{\sqrt{1-\gamma^{2}_{keb}}}{\gamma_{keb}} kau {\Bigg |} -\frac{\gamma^{2}_{keb}}{1-\gamma^{2}_{keb}} \right) \right],
\end{align}
where ${\rm sd}(\alpha|\nu)$ is the Jacobi elliptic function with real elliptic parameter $0\leqslant \nu \leqslant 1$~\cite{Abramowitz:1972,Lawden:1989}. Thus, constant $a$ can be written as
\begin{align}\label{cosa}
a=\frac{\gamma_{keb}}{k\sqrt{1-\gamma^{2}_{keb}}} 4 \bm K \left( -\frac{\gamma^{2}_{keb}}{1-\gamma^{2}_{keb}} \right) n, ~n\in Z^{+},
\end{align}
where $\bm K$ is the complete elliptic integral of the first kind which is the real quarter-period of the Jacobi elliptic function~\cite{Schwinger:1951nm}. We get a result from the Eq.~\eqref{weakfieldcondition}
\begin{align}
\frac{m\gamma_{keb}}{k\sqrt{1-\gamma^{2}_{keb}}} 2\pi n \gg 1.
\end{align}
It is a weak-field condition, and it must satisfies $E>B$ and $\sqrt{E^{2}-B^{2}}\ll E_{c}$ or $B_{c}$, as in the constant electromagnetic field case the Eq.~\eqref{1}.

It is easy to verify that the periodic instanton paths are
\begin{align}\label{pathcos1}
x_{1}&=\frac{i}{k}\frac{\gamma_{keb}}{\gamma_{kb}} {\rm arcsinh}\left[\frac{\gamma_{keb}}{\sqrt{1-\gamma^{2}_{keb}}}  ~{\rm cd}\left( \alpha(u) {\Bigg |} \nu \right)  \right],\\ \label{pathcos2}
x_{3}&=\frac{1}{k} {\rm arcsin}\left[ \frac{\gamma_{keb}}{\sqrt{1-\gamma^{2}_{keb}}}  ~{\rm sd}\left( \alpha(u) {\Bigg |} \nu \right) \right],\\ \label{pathcos3}
x_{4}&=\frac{1}{k}\frac{\gamma_{keb}}{\gamma_{ke}} {\rm arcsinh}\left[ \frac{\gamma_{keb}}{\sqrt{1-\gamma^{2}_{keb}}} ~{\rm cd}\left( \alpha(u) {\Bigg |} \nu \right) \right],
\end{align}
where $\alpha(u)=4n \bm K u$, $\bm K\equiv \bm K\left( \nu \right)$ and $\nu \equiv -\frac{\gamma^{2}_{keb}}{1-\gamma^{2}_{keb}}$.

\begin{figure}[ht!]\centering
\includegraphics[width=0.45\textwidth]{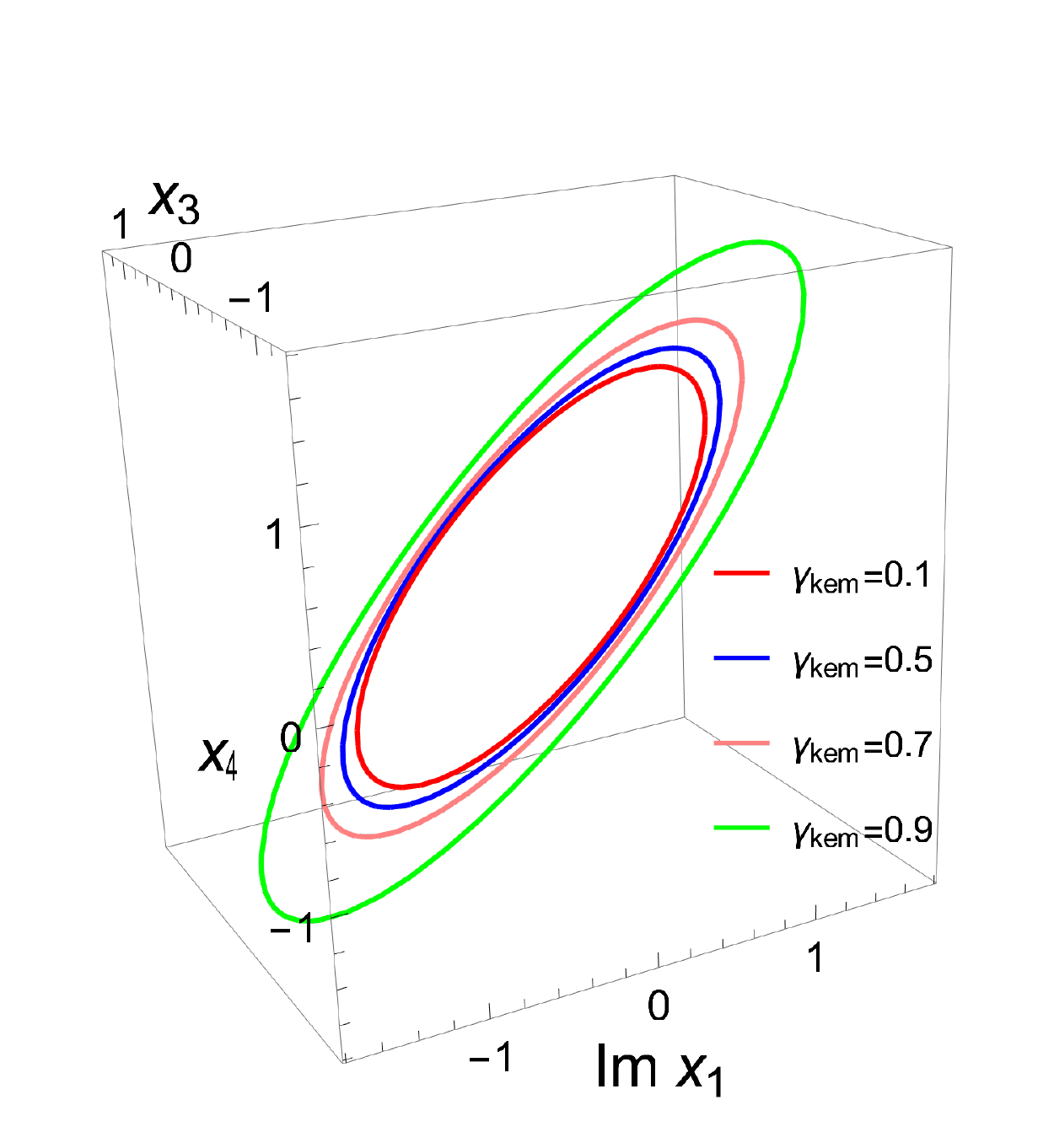}
\caption{Parametric plot of the stationary worldline instanton paths in the $({\rm Im}x_{1},~x_{3},~x_{4})$ space for the cases of the electric fields $E(x_{3})=E \cos(kx_{3})$ and the magnetic fields $B(x_{3})=B \cos(kx_{3})$ of strength $E$ and $B$. The paths are shown for various values of the parameter $\gamma_{keb}$(red, blue, pink and green for $0.1$, $0.5$, $0.7$ and $0.9$), ${\rm Im} x_{1}$, $x_{3}$ and $x_{4}$ have been expresses in units of $mB/e(E^{2}-B^{2})$, $m/e\sqrt{E^{2}-B^{2}}$ and $mE/e(E^{2}-B^{2})$.
\label{fig:cospath}}
\end{figure}

The stationary worldline instanton paths are the elliptic curves with the different $\gamma_{keb}$ as shown in Fig.~\ref{fig:cospath}. As $\gamma_{keb}$ increases, the instanton paths shrink in size. We can find that instanton loop paths tends to become the same elliptical shapes as shown in Fig.~\ref{fig:constantpaths} when $\gamma_{keb} \rightarrow 0$. However, the loop paths become infinitely large when $\gamma_{keb}=1$, and does not occur pair production. The reason of this phenomena is the same with in Sec.~\ref{sechfield}. Thus, there are pair creation only for $\gamma_{keb}<1$, and all of the elliptical shapes are in the same plane for any value of $\gamma_{keb}$. The stationary instanton paths have the the same result with Eq. (65) in Ref.~\cite{Dunne:2005sx} when $B=0$.

The stationary instanton action $\mathcal{S}_{0}$ can be written as
\begin{align}\label{actioncos}
\begin{split}
\mathcal{S}_{0}&=ma \int_{0}^{1}du ~{\rm cd}^{2}\left( \alpha(u) {\bigg |\nu} \right)\\
&=\frac{4nm}{k} \frac{\sqrt{1-\gamma^{2}_{keb}}}{\gamma_{keb}} \left( \bm E - \bm K \right)\\
&=\frac{4nm^{2}}{ \sqrt{E^{2}-B^{2}}} \frac{\sqrt{1-\gamma^{2}_{keb}}}{\gamma^{2}_{keb}} \left( \bm E - \bm K \right)\\
& \sim \begin{cases}
n\, \frac{m^2 \pi}{e \sqrt{E^{2}-B^{2}}}\left(1+\frac{\gamma_{keb}^{2}}{8}+\frac{3\gamma_{keb}^{4}}{64}+\dots\right), \quad \gamma_{keb}  \ll 1 \cr
n\,  \frac{4m^2 }{e \sqrt{E^{2}-B^{2}}}, \quad \gamma_{keb}\to 1,
\end{cases}
\end{split}
\end{align}
where $\bm E\equiv \bm E\left( \nu \right)$ is the complete elliptic integral of the second kind which is the real quarter-period of the Jacobi elliptic function~\cite{Abramowitz:1972,Lawden:1989}.
\begin{figure}[ht!]\centering
\includegraphics[width=0.45\textwidth]{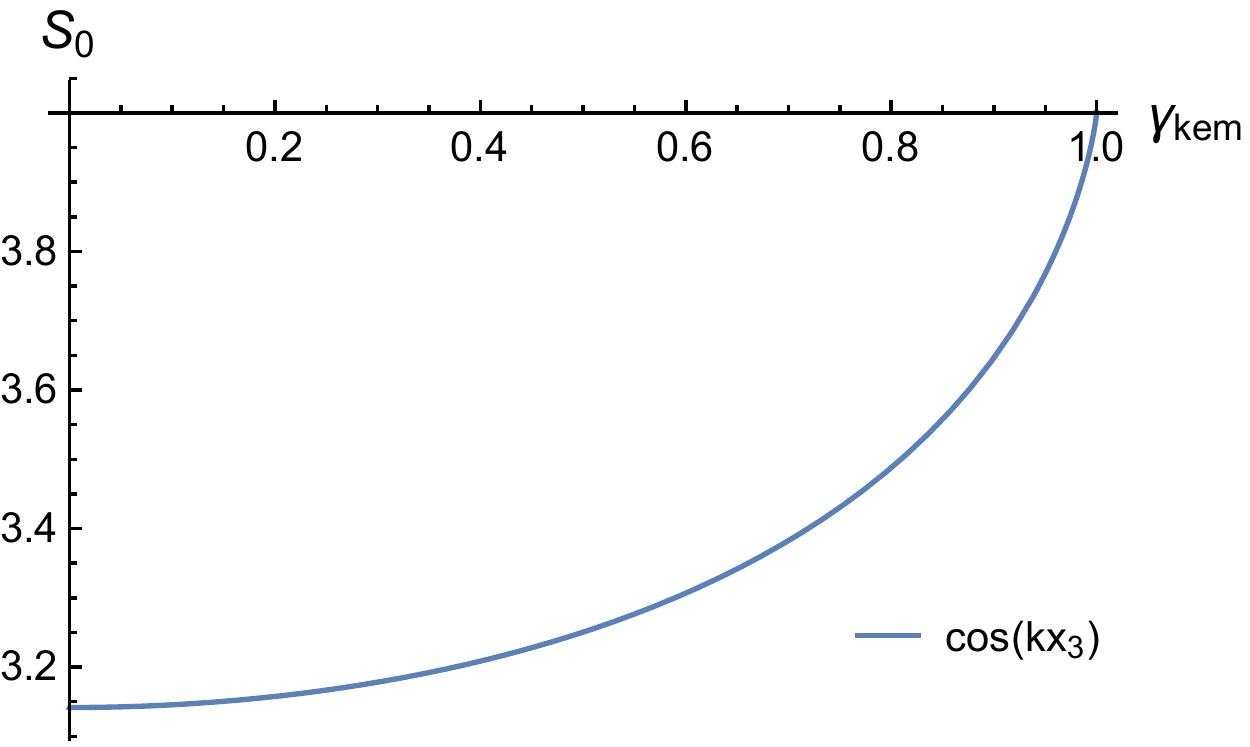}
\caption{Plot of the stationary worldline instanton action $S_{0}$ in the $({\rm Im}x_{1},~x_{3},~x_{3})$ space for the cases of the electric fields $E(x_{3})=E \cos(kx_{3})$ and the magnetic fields $B(x_{3})=B \cos(kx_{3})$ of strength E and B. The $S_{0}$ have been expresses in units of $nm^{2}/e\sqrt{(E^{2}-B^{2})}$.
\label{fig:cosaction}}
\end{figure}

This stationary instanton action $\mathcal{S}_{0}$ is plotted as a function of the $\gamma_{keb}$ in Fig.~\ref{fig:cosaction}. Note that the instantons action $\mathcal{S}_{0}$ increases with $\gamma_{keb}$, and $S_{0}=\pi$ and $4$ for $\gamma_{keb}=0$ and $1$, respectively. This result also is the same with in Ref.~\cite{Dunne:2005sx} when $B=0$.

In the next subsections, we study some electromagnetic background fields which are hard to solve analytically via Eqs.~\eqref{differentialequations1},~\eqref{differentialequations2} and \eqref{differentialequations3}. However, we can find the instanton action directly from Eq.~\eqref{worldline01} without calculation of instanton paths. It is very convenient way to predict the pair production probability and the pair-production time.
\subsection{Lorentzian field}
Now we study the electric field $E_{z}(x_{3})=E/(1+(kx_{3})^{2})^{3/2}$  and  magnetic field $B_{y}(x_{3})=B/(1+(kx_{3})^{2})^{3/2}$ in the Minkowski space, and  corresponding Euclidean space-dependent gauge potential is
\begin{align}\label{t2e}
A_{\mu}(x_{3})&=\left(-i\frac{E}{k}f(kx_{3}),~\frac{B}{k}f(kx_{3}),~0,~0\right),
\end{align}
where $f(kx_{3})=kx_{3}/(1+\left( kx_{3} \right)^{2})$ and $f'(kx_{3})=(1-\gamma^{2}_{keb} y )^{3/2}$, and then we can obtain the instanton action $\mathcal{S}_{0}$ directly from Eq.~\eqref{worldline01}
\begin{align}\label{actiont2}
\begin{split}
\mathcal{S}_{0}&=\frac{4mn}{k\gamma_{keb}} \int_{0}^{1}dy \frac{\sqrt{1-y^{2}}}{|f^{'}|}\\
&=\frac{4nm^{2}}{ \sqrt{E^{2}-B^{2}}} \frac{1}{\gamma^{2}_{keb}} \left( \bm K \left( \gamma^{2}_{keb} \right) - \bm E \left( \gamma^{2}_{keb} \right) \right).
\end{split}
\end{align}
\begin{figure}[ht!]\centering
\includegraphics[width=0.45\textwidth]{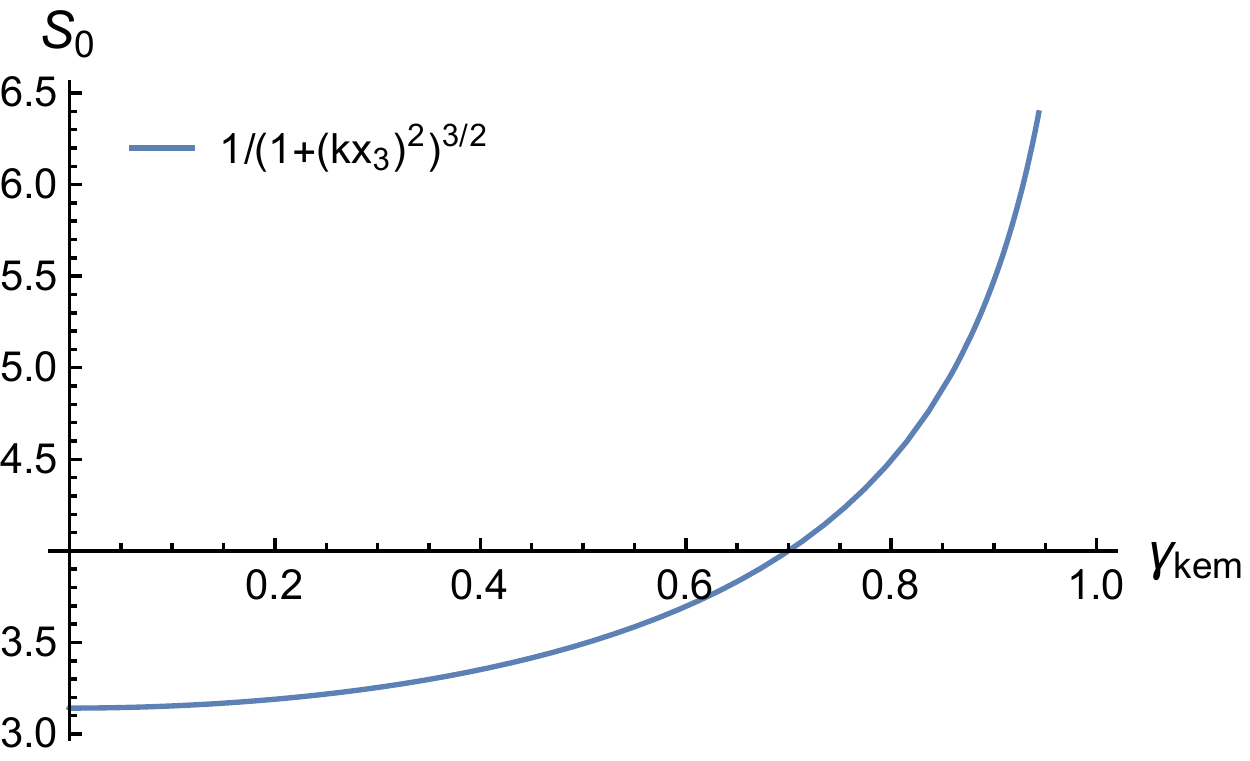}
\caption{Plot of the stationary worldline instanton action $S_{0}$ in the $({\rm Im}x_{1},~x_{3},~x_{3})$ space for the cases of the electric fields $E(x_{3})= E/(1+(kx_{3})^{2})^{3/2}$ and the magnetic fields $B(x_{3})= B/(1+(kx_{3})^{2})^{3/2}$ of strength $E$ and $B$, in units of $nm^{2}/e\sqrt{(E^{2}-B^{2})}$, plotted as a function of the various parameter values of $\gamma_{keb}$.
\label{fig:lorentzaction}}
\end{figure}
The instanton action $S_{0}=\pi$ and $2\pi$ for $\gamma_{keb}=0$ and $1$, respectively. We can find that the instantons action increases and pair creation rate decreases with $\gamma_{keb}$. It is similar to results in Sec.~\ref{sechfield} and Sec.~\ref{cosfield}. We can obtain the same result of instanton action $\mathcal{S}_{0}$  in Ref.~\cite{Dunne:2005sx} when $B=0$.
\subsection{Elliptic cosine field}
\begin{figure}[ht!]\centering
\includegraphics[width=0.45\textwidth]{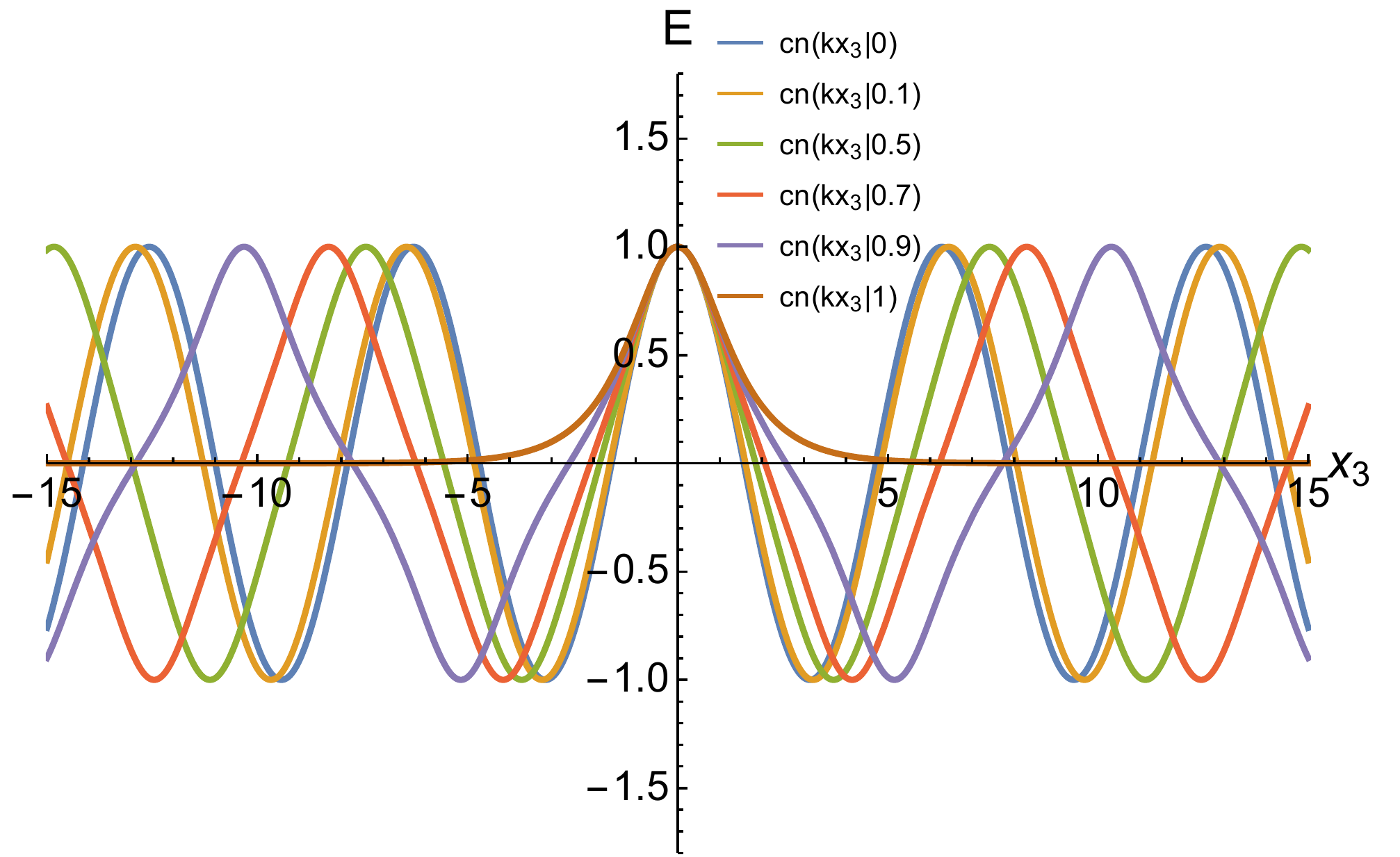}
\caption{Plot of the space-dependent electromagnetic fields $E_{z}(x_{3})=E {\rm cn}(kx_{3}|q)$ and $B_{y}(x_{3})=B {\rm cn}(kx_{3} | q )$ for different values of the parameter $q$. The distributions for blue, yellow, green, magenta, lightblue and orange in the figure corresponds to $0$, $0.1$, $0.5$, $0.7$, $0.9$ and $1$ of $q$.
\label{fig:cnfield}}
\end{figure}

\begin{figure}[ht!]\centering
\includegraphics[width=0.45\textwidth]{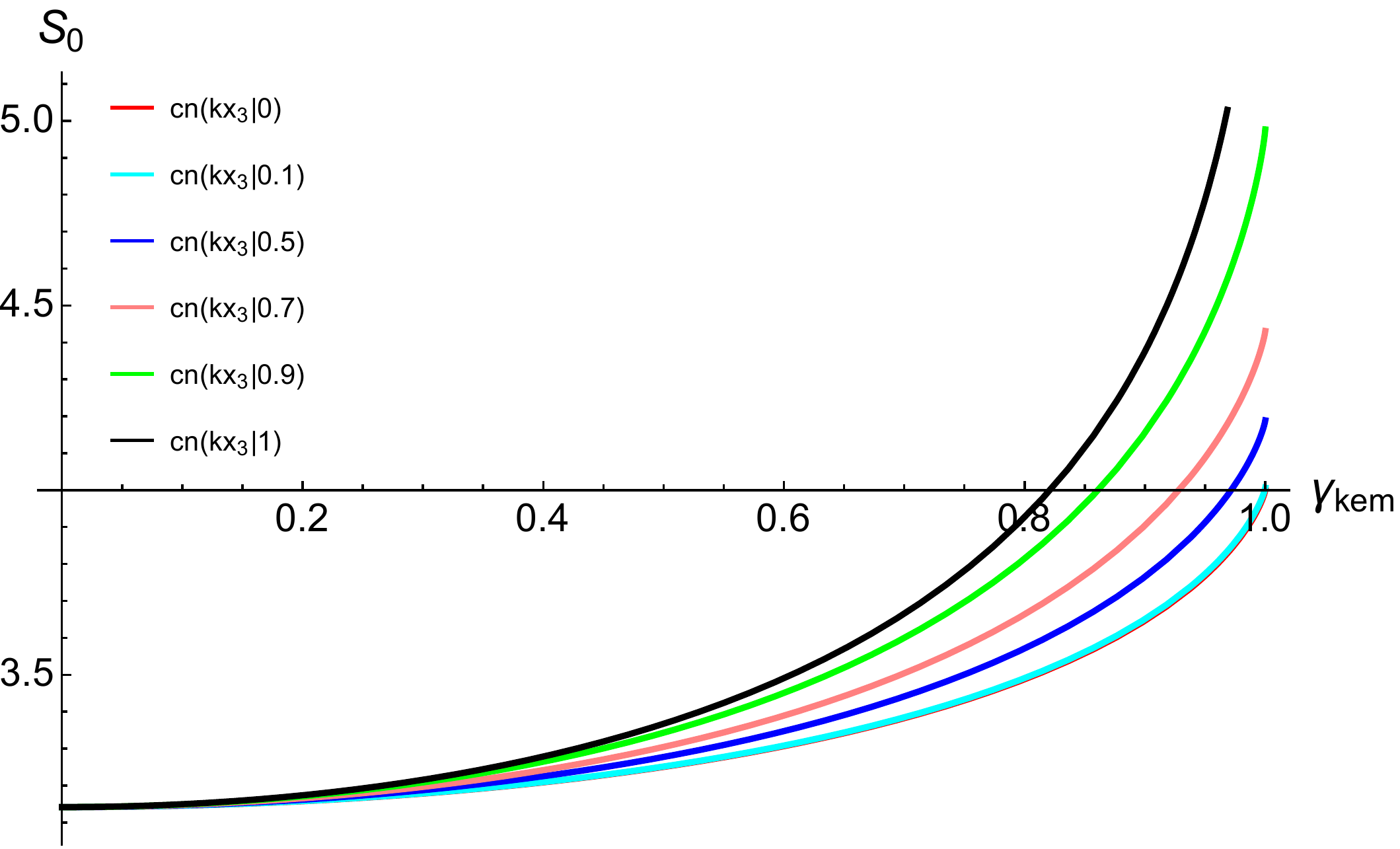}
\caption{Plot of the stationary worldline instanton action $S_{0}$ in the $({\rm Im}x_{1},~x_{3},~x_{3})$ space for the cases the spatially inhomogeneous electromagnetic fields $E_{z}(x_{3})=E {\rm cn}(kx_{3}|q)$ and $B_{y}(x_{3})=B {\rm cn}(kx_{3} | q )$ of strength $E$ and $B$ with different values of the parameter $q$. The distributions for blue, yellow, green, magenta, lightblue and orange in the figure corresponds to $0$, $0.1$, $0.5$, $0.7$, $0.9$ and $1$ of $q$. The $S_{0}$ have been expresses in units of $nm^{2}/e\sqrt{(E^{2}-B^{2})}$.
\label{fig:allaction}}
\end{figure}

In this subsection, we discuss the more complicated electric field $E_{z}(x_{3})=E {\rm cn}(kx_{3} | q )$  and  magnetic field $B_{y}(x_{3})=B {\rm cn}(kx_{3} | q )$ for various values of parameter $q$ in the Minkowski space as shown in Fig.~\ref{fig:cnfield}. Corresponding Euclidean space-dependent gauge potential is
\begin{align}\label{cneb}
A_{\mu}(x_{3})&=\left(-i\frac{E}{k}f(kx_{3}),~\frac{B}{k}f(kx_{3}),~0,~0\right),
\end{align}
where $f(kx_{3})={\rm sn}(\upsilon |q)$, $f'=\sqrt{1-\gamma^{2}_{keb} y^{2}}\sqrt{1-q^{2}\gamma^{2}_{keb} y^{2}}$, ${\rm sn}(\upsilon |q)$ and ${\rm cn}(\upsilon |q)$ are the Jacobi elliptic sine and cosine functions~\cite{Abramowitz:1972,Lawden:1989}, the two variables $\upsilon$ and $q$ in the elliptic functions denote the terms of the amplitude and elliptic modulus. Note that ${\rm cn}(kx_{3}|0)=\cos(kx_{3})$ and ${\rm cn}(kx_{3}|1)={\rm sech}(kx_{3})$, because the ${\rm cn}(kx_{3}|q)$ contains an infinite number of functions as shown in Fig.~\ref{fig:cnfield}. We can obtain the instanton action $\mathcal{S}_{0}$ from Eq.~\eqref{worldline01} directly
\begin{align}\label{actiont3}
\begin{split}
\mathcal{S}_{0}&=\frac{4mn}{k\gamma_{keb}} \int_{0}^{1}dy \frac{\sqrt{1-y^{2}}}{|f^{'}|}=\frac{4nm^{2}}{ \sqrt{E^{2}-B^{2}}}\frac{1}{\sqrt{\gamma^{2}_{keb}-1}} \\
&\bm \prod \left( \frac{1}{\sqrt{\gamma^{2}_{keb}-1}}, {\rm arcsin}\left( \frac{y\sqrt{\gamma^{2}_{keb}-1}}{\sqrt{1-y^{2}}}\right), \frac{1-q\gamma^{2}_{keb}}{1-\gamma^{2}_{keb}}  \right) {\Bigg |}_{0}^{1},
\end{split}
\end{align}
where $\bm \prod \left( \alpha, \phi, \beta  \right)$ is the incomplete elliptic integral of the third kind in which have three parameters independently~\cite{Abramowitz:1972,Lawden:1989}
\begin{align}
\bm \prod \left( \alpha, \phi, \beta  \right)&=\int_{0}^{\phi}\frac{1}{1-\alpha {\rm sin}^{2}\theta} \frac{d\theta}{\sqrt{1- \left( {\rm sin} \theta {\rm sin} \beta \right)}}.
\end{align}
If $q=0$, the instanton action can be written as
\begin{align}
\begin{split}
\mathcal{S}_{0}&=\frac{4nm^{2}}{ \sqrt{E^{2}-B^{2}}}  \frac{\left( \bm E \left( \gamma^{2}_{keb} \right) - (1-\gamma^{2}_{keb}) \bm K \left( \gamma^{2}_{keb} \right) \right)}{\gamma^{2}_{keb}}\\
&\equiv \frac{4nm^{2}}{ \sqrt{E^{2}-B^{2}}}  \left( \bm E \left( \frac{-\gamma^{2}_{keb}}{1-\gamma^{2}_{keb}} \right) -  \bm K \left( \frac{-\gamma^{2}_{keb}}{1-\gamma^{2}_{keb}} \right) \right)\\
&~~~~~ \times \frac{\sqrt{1-\gamma^{2}_{keb}}}{\gamma^{2}_{keb}}.
\end{split}
\end{align}
It is the same result with the Eq.~\eqref{actioncos} for $q=0$, and the same approximate result with the Eq.~\eqref{actionsech} for $q=1$, as shown in Fig.~\ref{fig:allaction} in units of $nm^{2}/e\sqrt{(E^{2}-B^{2})}$.

We can find that the instantons action increases and the pair production rate decreases with $\gamma_{keb}$. The instantons action equals to $\pi$ and $4$ for $\gamma_{keb}= 0$ and $\gamma_{keb}= 1$ when $q=0$, and $\pi$ and $2\pi$ for $\gamma_{keb}= 0$ and $\gamma_{keb}= 1$ when $q=1$. The instantons action values for other fields between $0\leqslant q \leqslant 1$ are within the region from $\pi$ to $2\pi$, as shown in Fig.~\ref{fig:allaction}. This is the same result with in Sec.~\ref{sechfield} and Sec.~\ref{cosfield}. We can obtain the same value of instanton action of $\mathcal{S}_{0}$ as $B=0$ in Ref.~\cite{Dunne:2005sx}.
\section{pair-production time}\label{tunnelingtime}
\subsection{Vacuum-decay time}
In this section, we discuss the vacuum-decay time (include tunneling process, multiphoton absorption process) for space-dependent electromagnetic fields analytically. Because the vacuum decay in our cases belongs to the energy transformation process from field to particles, from Eq.~(6) in Ref.~\cite{Labun:2008re} we can obtain the total pair-production time as
\begin{align}\label{tunnelingtime11}
\begin{split}
\mathcal{T}_{d}&= \mathcal{U} _f\left(\ \frac{d \langle \mathcal{U}_m\rangle }{dt} \right)^{-1},
\end{split}
\end{align}
where $\mathcal{T}_{d}$ denotes the vacuum-decay time, $\mathcal{U} _m$ and $\mathcal{U} _f$ are the transverse energy density and the field energy density. This energy density of the field is expressed in terms of the invariants $\mathcal{F}=(1/4)F^{\mu\nu}F_{\mu\nu}=\frac{1}{2}(B^2-E^2)$ and $\mathcal{G}=(1/4)F^{\mu\nu}F^*_{\mu\nu}=E\cdot B$ as \cite{Labun:2008re,Labun:2008qq}
\begin{align}\label{uf}
\mathcal{U} _f&= \sqrt {\left(\mathcal{F}^2+\mathcal{G}^2\right)f^2\left(\mathcal{F},\mathcal{G}\right)+A^2\left(\mathcal{F},\mathcal{G}\right)},\\ \label{l}
\mathcal{L} &= - \mathcal{F} + \frac{8\alpha^{2}}{45 m^{4}} \mathcal{F}^{2} + \frac{14\alpha^{2}}{45 m^{4}} \mathcal{G}^{2},
\end{align}
where $\mathcal{L}$ is the effective Heisenberg-Euler Lagrangian, $f=\partial \mathcal{L}/\partial \mathcal{F} \to -1$ for the Maxwell's equation when the lowest-order approximation is taken into account, $A$ is the conformal anomaly induced by external fields, $T^\mu_\mu =4A$~\cite{Labun:2008re}.

\begin{figure}[ht!]\centering
\includegraphics[width=0.45\textwidth]{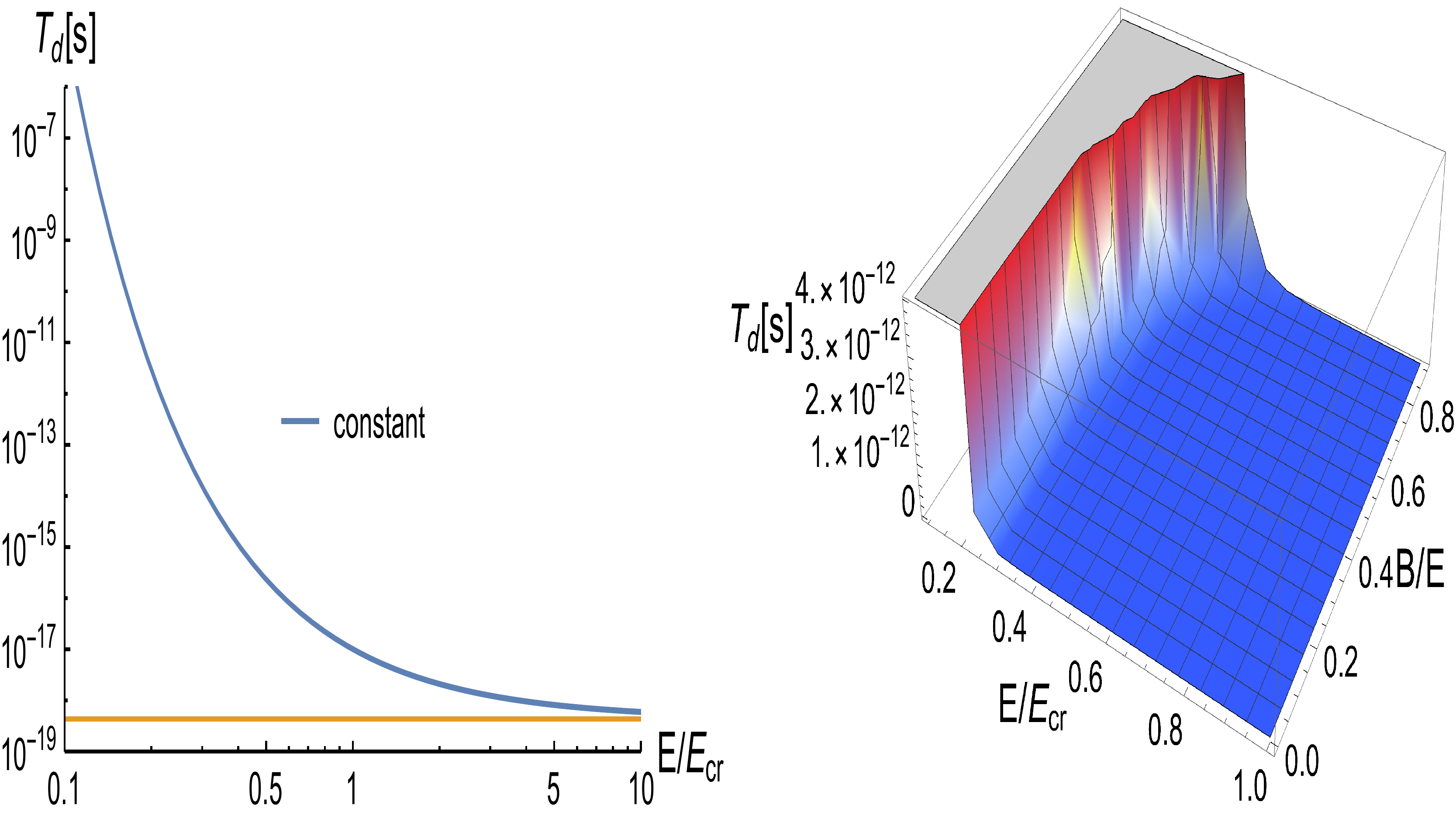}
\caption{Plot of the vacuum-decay time for constant electric field (left) for normalized field $E/E_{cr}$ and constant electromagnetic field (right) for ratios of $E/E_{cr}$ and $B/E$.
\label{fig:tunnelingtimeconstant}}
\end{figure}

If we redefine the rate of vacuum decay (combine with Eqs.~(2) and (4) in Ref.~\cite{Labun:2008re}) by replacing the worldline instantons action $\mathcal{S}_{0}$ with the action under constant electric field as
\begin{align}\label{rateofvacuum}
\begin{split}
\frac{d \langle \mathcal{U}_m\rangle }{dt}&= \frac{eE}{2\pi^2}\int_m^{\infty} d\epsilon _{\perp} \,2\epsilon^2_{\perp} e^{-\beta(\gamma_{keb})\epsilon_{\perp}^2}\\
&=\frac{eE}{4 \pi^{2}} \frac{m e^{- \mathcal{S}_{0}}}{\beta(\gamma_{keb})} \left( 1+\frac{\sqrt{\pi} e^{ \mathcal{S}_{0}} }{2\sqrt{\mathcal{S}_{0}}} \rm{efrc} \left( \sqrt{\mathcal{S}_{0}} \right) \right),
\end{split}
\end{align}
where $\beta(\gamma_{keb})=\mathcal{S}_{0}/m^{2}$, $\epsilon _{\perp}=\sqrt{m^2+p^{2}_{\perp}}$. Therefore, the vacuum-decay time can be obtained as
\begin{align}\label{tunnelingtime2}
\begin{split}
\mathcal{T}_{d}&= \mathcal{U} _f \left( \frac{eE}{4 \pi^{2}} \frac{m e^{- \mathcal{S}_{0}}}{\beta(\gamma_{keb})} \left( 1+\frac{\sqrt{\pi} e^{ \mathcal{S}_{0}} }{2\sqrt{\mathcal{S}_{0}}} \rm{efrc} \left( \sqrt{\mathcal{S}_{0}} \right) \right) \right)^{-1}.
\end{split}
\end{align}
If we consider constant electric field, the vacuum-decay time becomes that
\begin{align}\label{tunnelingtime3}
\begin{split}
\mathcal{T}_{d}&= \mathcal{U} _f \left( \omega_0 E^2e^{\frac{-\pi E_0}{E}}
\left\{ 1+ h\!\left(\!\sqrt{\frac{\pi E_0}{E}}\right) \right\}  \right)^{-1},
\end{split}
\end{align}
where $\omega_{0} = \alpha c/ \pi^2 \lambdabar_{e}=\alpha mc^2/ \pi^2\hbar=5.740\times 10^{17}$s$^{-1}$ and $h(z) =  {\sqrt{\pi}}e^{z^2}\mathrm{erfc}(z)/{2z}$. If we ignore the second and third terms of $\mathcal{L}$ in Eq.~\eqref{l}, it is the same result with Eq.~(6) in Ref.~\cite{Labun:2008re} when $B/E=0$ (see Fig.~\ref{fig:tunnelingtimeconstant}). The orange line in Fig.~\ref{fig:tunnelingtimeconstant} (left) represents the minimum value ($\omega^{-1}_{0}/4$) of the vacuum-decay time for constant electric field. We can find that the time increases with the decrease of $\gamma_{keb}$ for the same parameter $B/E$, but does not depend on the $x_{3}$.

\begin{figure}[ht!]\centering
\includegraphics[width=0.44\textwidth]{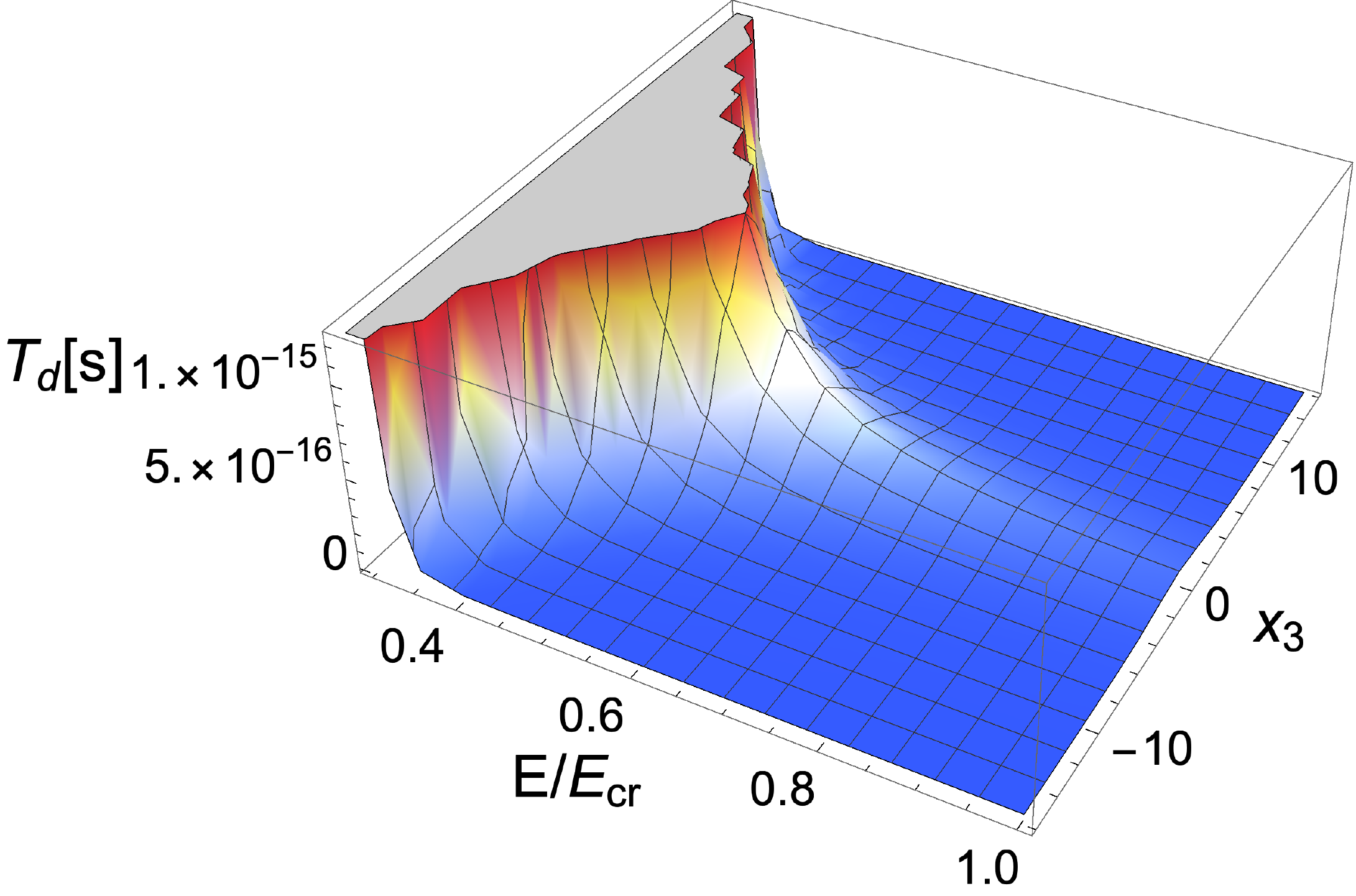}
\includegraphics[width=0.44\textwidth]{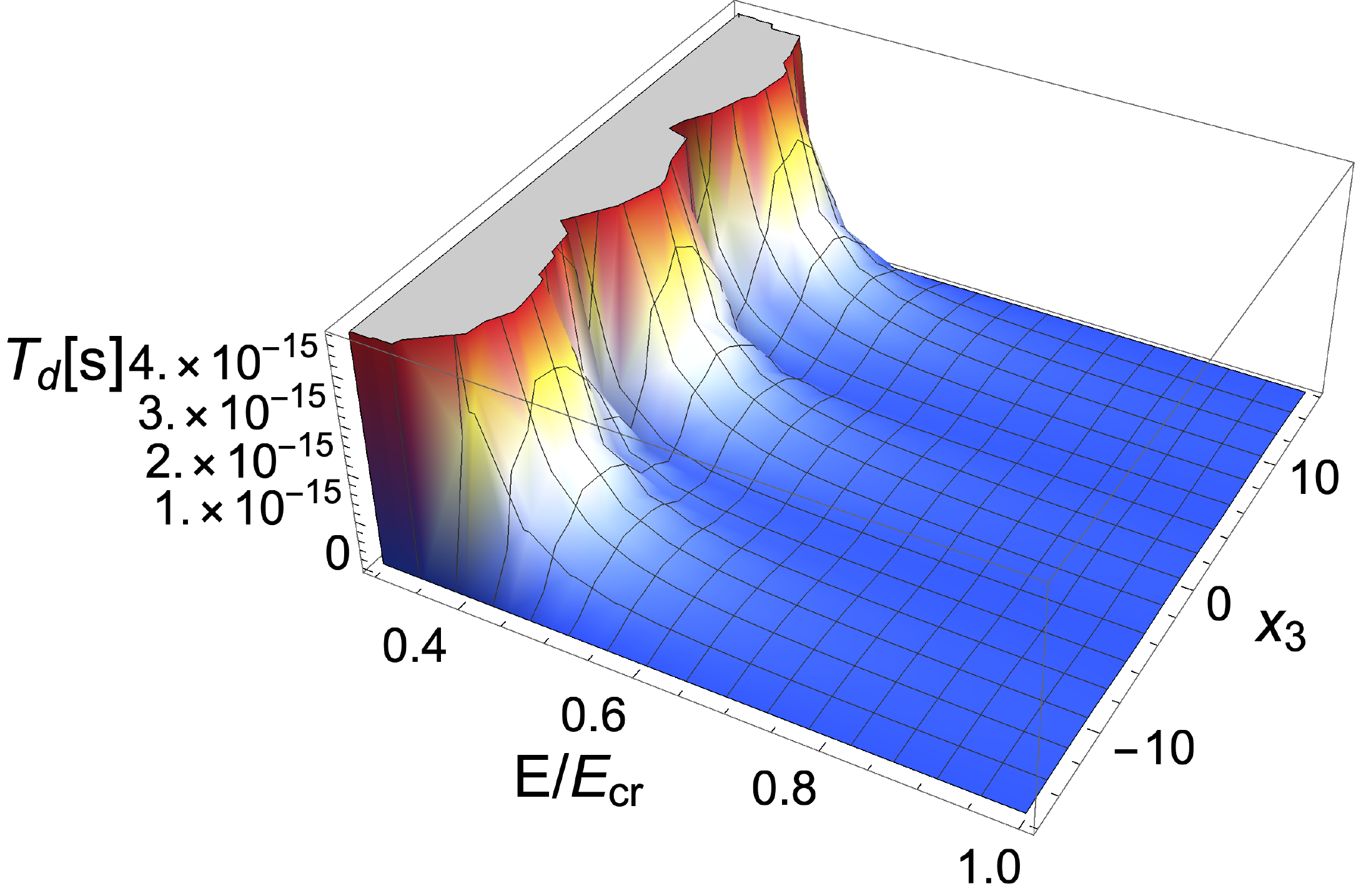}
\includegraphics[width=0.44\textwidth]{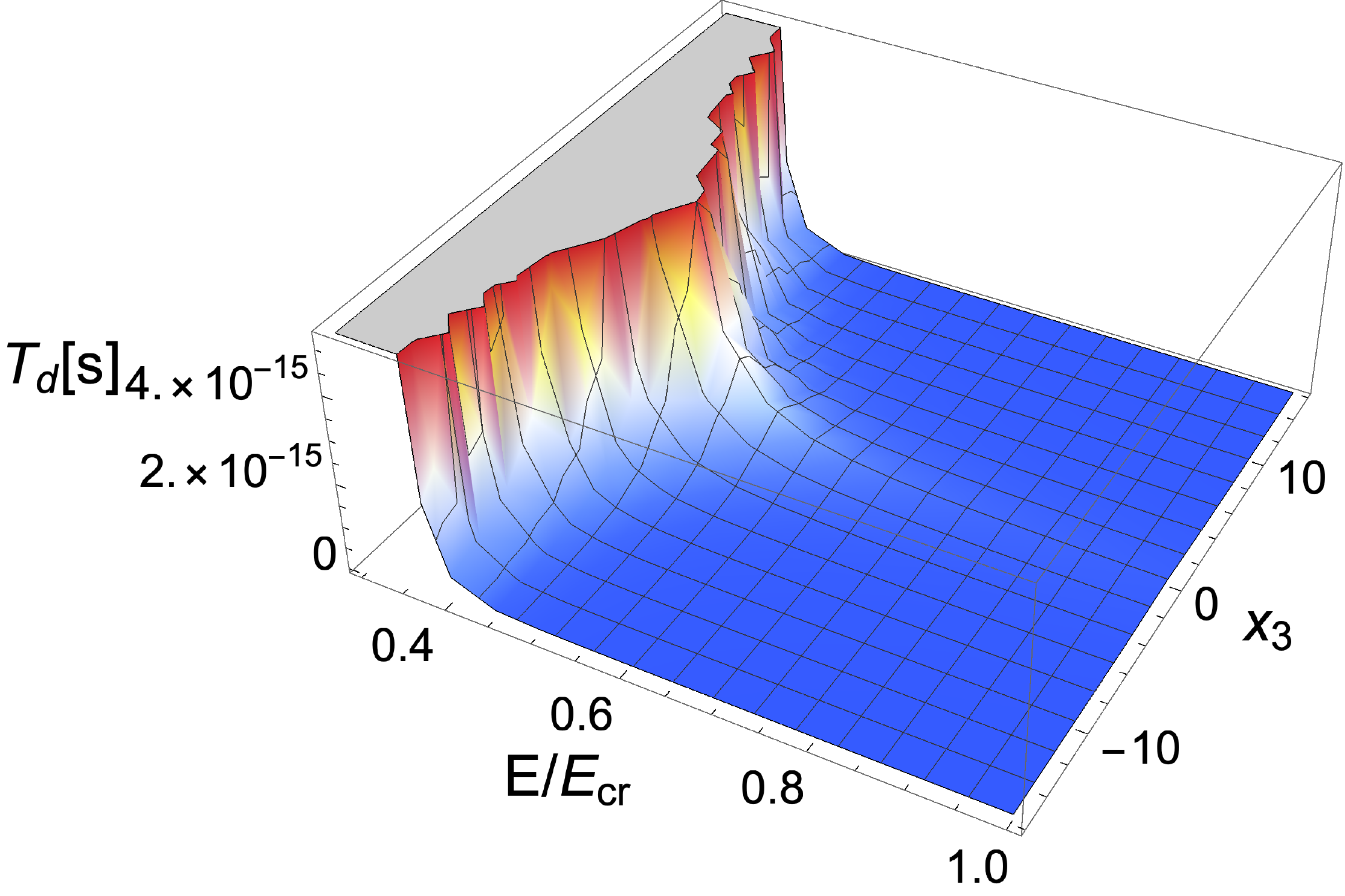}
\includegraphics[width=0.44\textwidth]{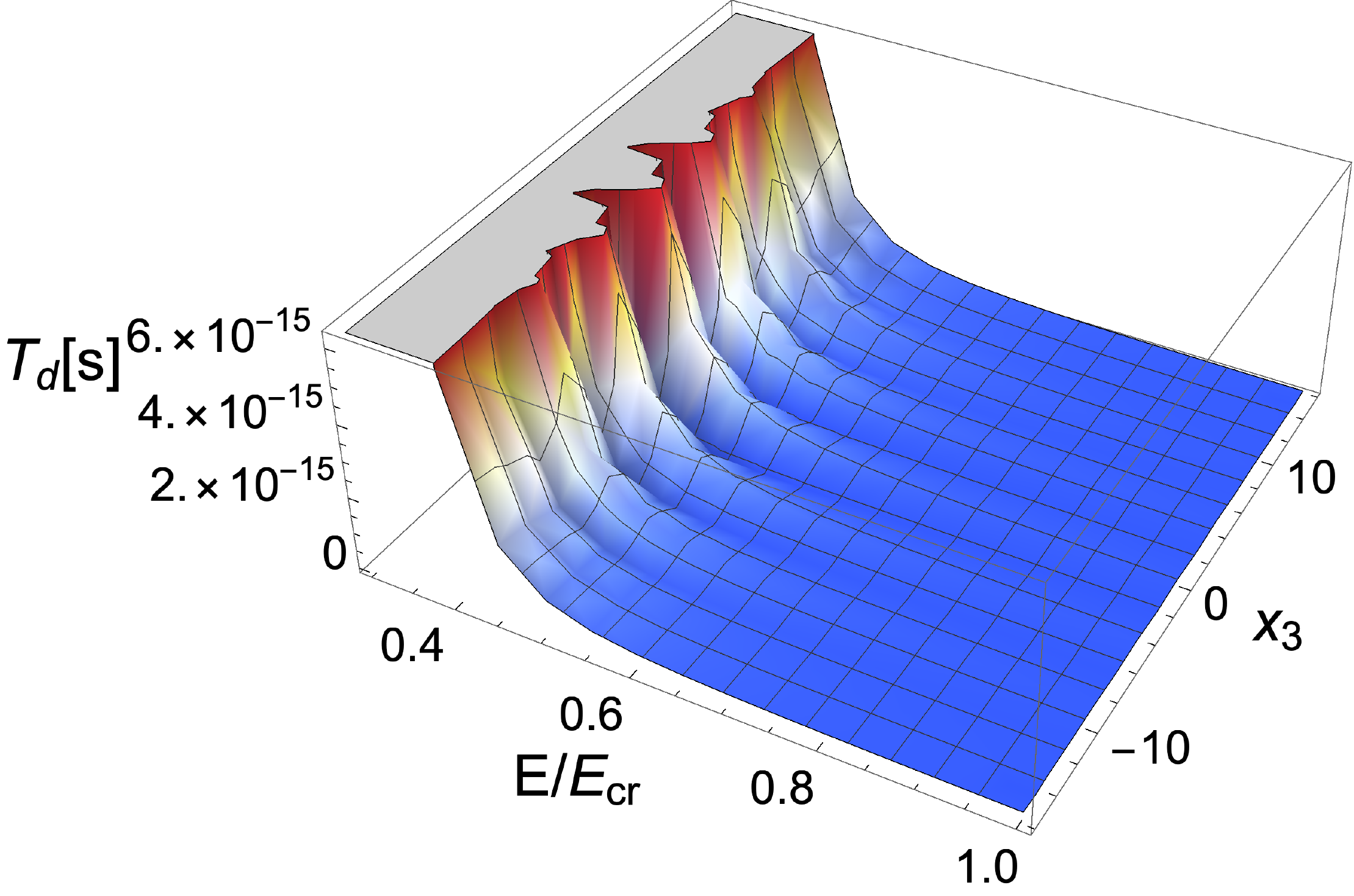}
\caption{Plot of the vacuum-decay time for ${\rm sech}^{2}\left( kx_{3} \right)$, ${\rm cos}\left( kx_{3} \right)$, $1/\left( 1 + (k x_{3})^{2} \right)^{3/2}$ and ${\rm cn}\left( k x_{3} \vert 0.7 \right)$ with $k=0.3$ and $E/B=0.01$ from top to bottom with ratios of $E/E_{cr}$ and $x_{3}$.
\label{fig:tunnelingtimeall}}
\end{figure}

From Eq.~\eqref{tunnelingtime2}, we can find that the vacuum-decay time depends on $x_{3}$ and electromagnetic field modes. To facilitate the discussion of the effect of coordinate space $x_{3}$ and electromagnetic field modes, we give four examples of different electromagnetic fields with parameters $k=0.3$ and $E/B=0.01$ for the Maxwell's equation as shown in Fig.~\ref{fig:tunnelingtimeall}. We can find that the time increases with decrease of parameter $\gamma_{keb}$ when $x_3$ is fixed. Meanwhile, if we consider the overall effect in the distributions of vacuum-decay time for ${\rm sech}^{2}\left( kx_{3} \right)$, ${\rm cos}\left( kx_{3} \right)$, $1/\left( 1 + (k x_{3})^{2} \right)^{3/2}$ and ${\rm cn}\left( k x_{3} \vert 0.7 \right)$, the distributions of these times are similar to those shapes of the applied electromagnetic field modes for the same parameter $E/E_{cr}$.

Certainly, we can apply Eq.~\eqref{tunnelingtime2} not only to higher-order Heisenberg-Euler Lagrangian, but also to multiphoton and tunneling processes (or mixing process).
\subsection{Tunneling time}
In this subsection, we discuss the expression of the tunneling time of the Bohm viewpoint by using the instanton action. From Ref.~\cite{Landauer:1994zz}, the tunneling time can be written by using  Eq.~\eqref{tunnelingtimeappendixb7} in Appendix~\ref{app:Bppa}
\begin{align}\label{tunnelingtimebohm}
\mathcal{T}_{t}&=\int_{u\left(x^{-}\right)}^{u\left(x^{+}\right)}  \dot{x}_{4}  du=\int_{x_{3}^{-}}^{x_{3}^{+}} \frac{ \dot{x}_{4} }{\dot{x}_{3}} dx_{3},
\end{align}
where $\mathcal{T}_{t}$ denotes the tunneling time. According to Eqs.~\eqref{differentialequations21} and \eqref{differentialequations31}, we can obtain that
\begin{align}\label{tunnelingtimebohm1}
\begin{split}
\mathcal{T}_{t}&=\int_{x_{3}^{-}}^{x_{3}^{+}} \frac{ - \frac{aeE}{mk}  f\left(kx_{3}\right)}{ a \sqrt{1- \frac{f^{2} \left(kx_{3}\right)}{\gamma^{2}_{keb}}} } dx_{3}\\
&=-\frac{2}{\gamma_{ke}} \int_{0}^{x_{3}^{+}} \frac{ f\left(kx_{3}\right)}{ \sqrt{1- \frac{f^{2} \left(kx_{3}\right)}{\gamma^{2}_{keb}}} } dx_{3}.
\end{split}
\end{align}
Thus the tunneling times corresponding to the space-dependent electromagnetic fields researched in Sec.~\ref{model} are
\begin{align}\label{tunnelingtimebohmall}
\begin{split}
\mathcal{T}_{t}&=\begin{cases}
\frac{2\gamma^{2}_{keb}}{\gamma_{ke}}, ~~~~~~~~~~~~~~~~~~~~~~~~~~~~~~~~~~~~~~~~~{\rm constant} \\
\frac{2}{k}\frac{\gamma_{keb}}{\gamma_{ke}\sqrt{1-\gamma^{2}_{keb}}} {\rm arcsin}\left[ \gamma_{keb} \right],~~~~~~~~~~{\rm sech}^{2}\left(k x_{3}\right) \\
\frac{2}{k}\frac{\gamma_{keb}}{\gamma_{ke}} {\rm arcsinh}\left[ \frac{\gamma_{keb}}{\sqrt{1-\gamma^{2}_{keb}}} \right], ~~~~~~~~~~~~{\rm cos}\left(k x_{3}\right)\\
-\frac{2}{\gamma_{keb}} \int_{0}^{x_{3}^{+}} \frac{\frac{kx_{3}}{\left(1+\left(kx_{3}\right)^{2}\right)}}{\sqrt{1-\frac{\left(\frac{kx_{3}}{\left(1+\left(kx_{3}\right)^{2}\right)}\right)^{2}}{\gamma_{keb}^{2}}}} dx_{3}, ~~~\frac{1}{(1+(kx_{3})^{2})^{3/2}}\\
-\frac{2}{\gamma_{keb}} \int_{0}^{x_{3}^{+}} \frac{{\rm sn}\left(kx_{3} | q \right)}{\sqrt{1-\frac{{\rm sn}^{2}\left(kx_{3} | q \right)}{\gamma_{keb}^{2}}}} dx_{3} , ~~~~~~~~~{\rm cn}\left(kx_{3} | q \right)
\end{cases}
\end{split}
\end{align}
respectively. On the other hand, we know that there is a simple relationship between the tunneling time $\mathcal{T}_{t}$ and the $x_{4}$ as
\begin{align}\label{tunnelingtimea}
\mathcal{T}_{t}=2 x_{4}\left( 0 \right)= \left( x_{4}^{max} - x_{4}^{min}\right).
\end{align}
For example, it is valid for cases of constant, ${\rm sech}^{2}\left(kx_{3}\right)$ and ${\rm cos}\left(kx_{3}\right)$ fields by compering Eqs.~\eqref{patheb3}, \eqref{pathsech3}, \eqref{pathcos3} with  \eqref{tunnelingtimebohmall} where the instanton number $n=1$ is given.
Obviously the relationship Eq.\eqref{tunnelingtimea} is valid generally for any field that has instanton solutions, which is proven in Appendix~\ref{app:Bppa} by using different methods. Physically, the tunneling time is just one that an instanton "particle" takes when it travels the half loop path from $x^{-}_{3}$ to $x^{+}_{3}$, see Figs.~\ref{fig:constantpaths}, ~\ref{fig:sechpath} and ~\ref{fig:cospath}.

We can find another nature of tunneling time for the worldline instanton approach that according to Eq.~\eqref{tunnelingtimebohmall} the tunneling time increases with parameter $\gamma_{keb}$ when $\gamma_{ke}$ and $k$ are given, while the pair production rate decreases. Thus, the larger the tunneling time, the smaller the pair creation. Interestingly, the tunneling time no longer depends on the $x_{3}$, which is different from the vacuum decay time.
\section{Discussion}\label{discussion}
\begin{figure*}[ht!]\centering
\includegraphics[width=0.45\textwidth]{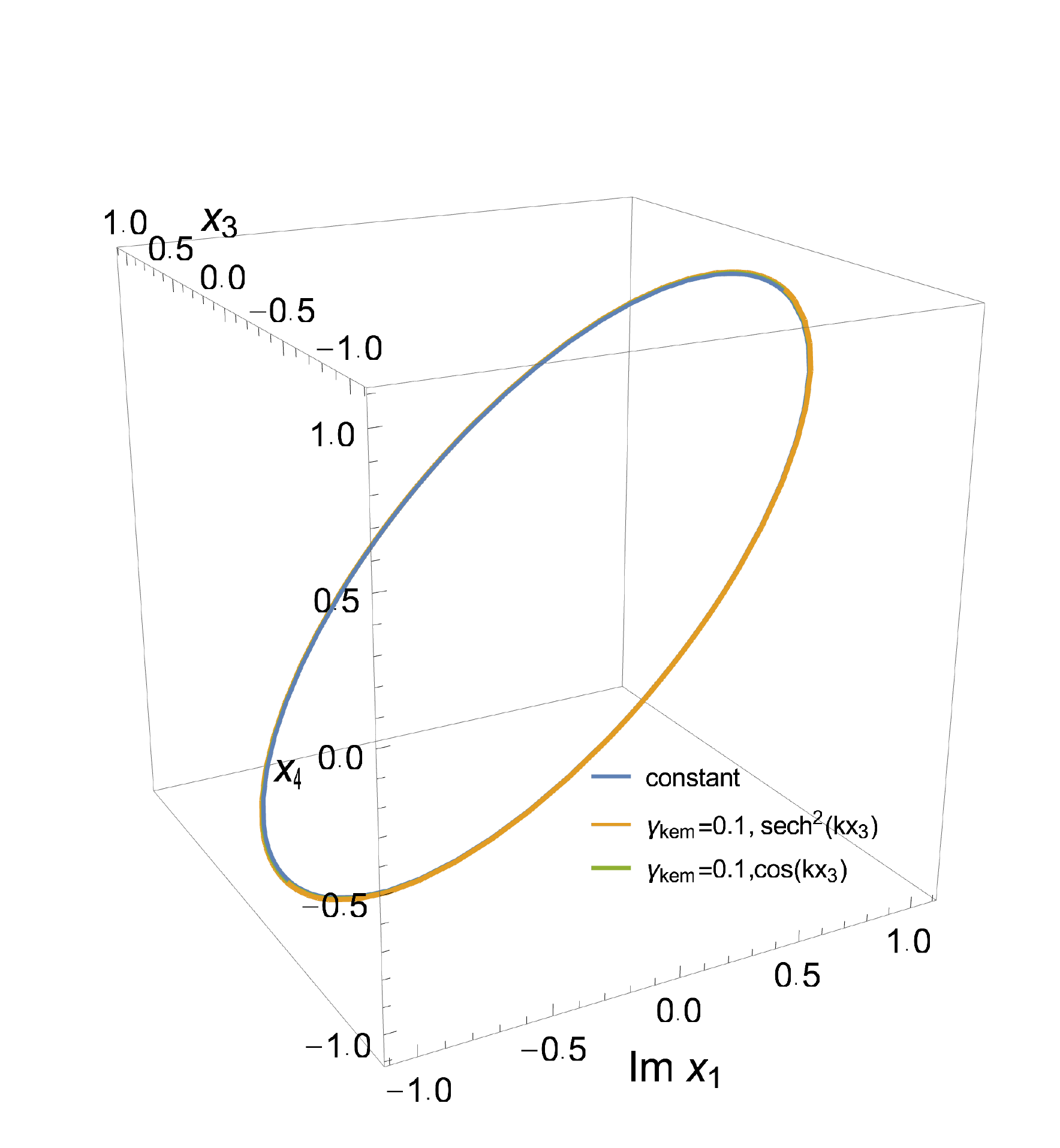}
\includegraphics[width=0.45\textwidth]{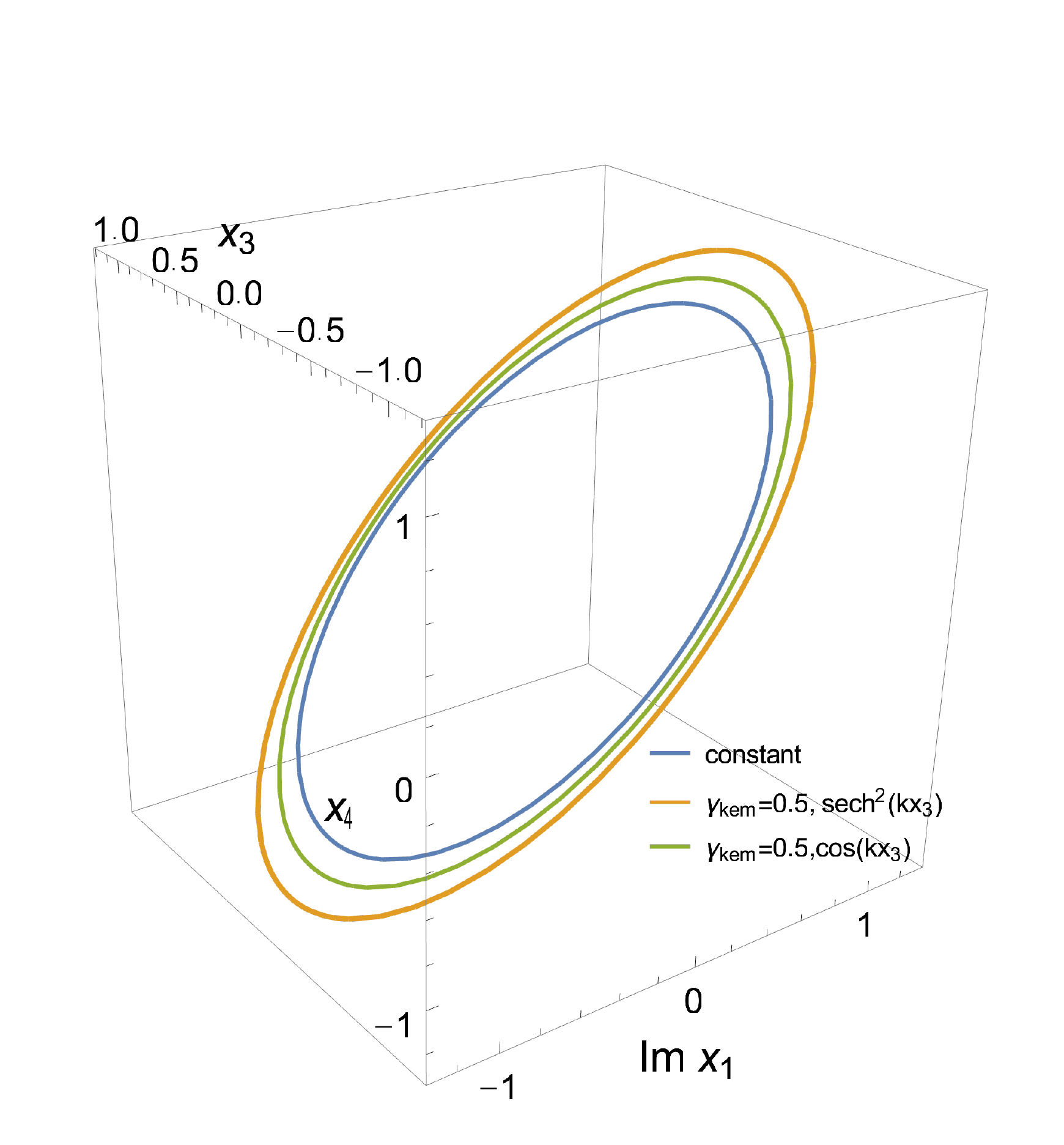}
\includegraphics[width=0.45\textwidth]{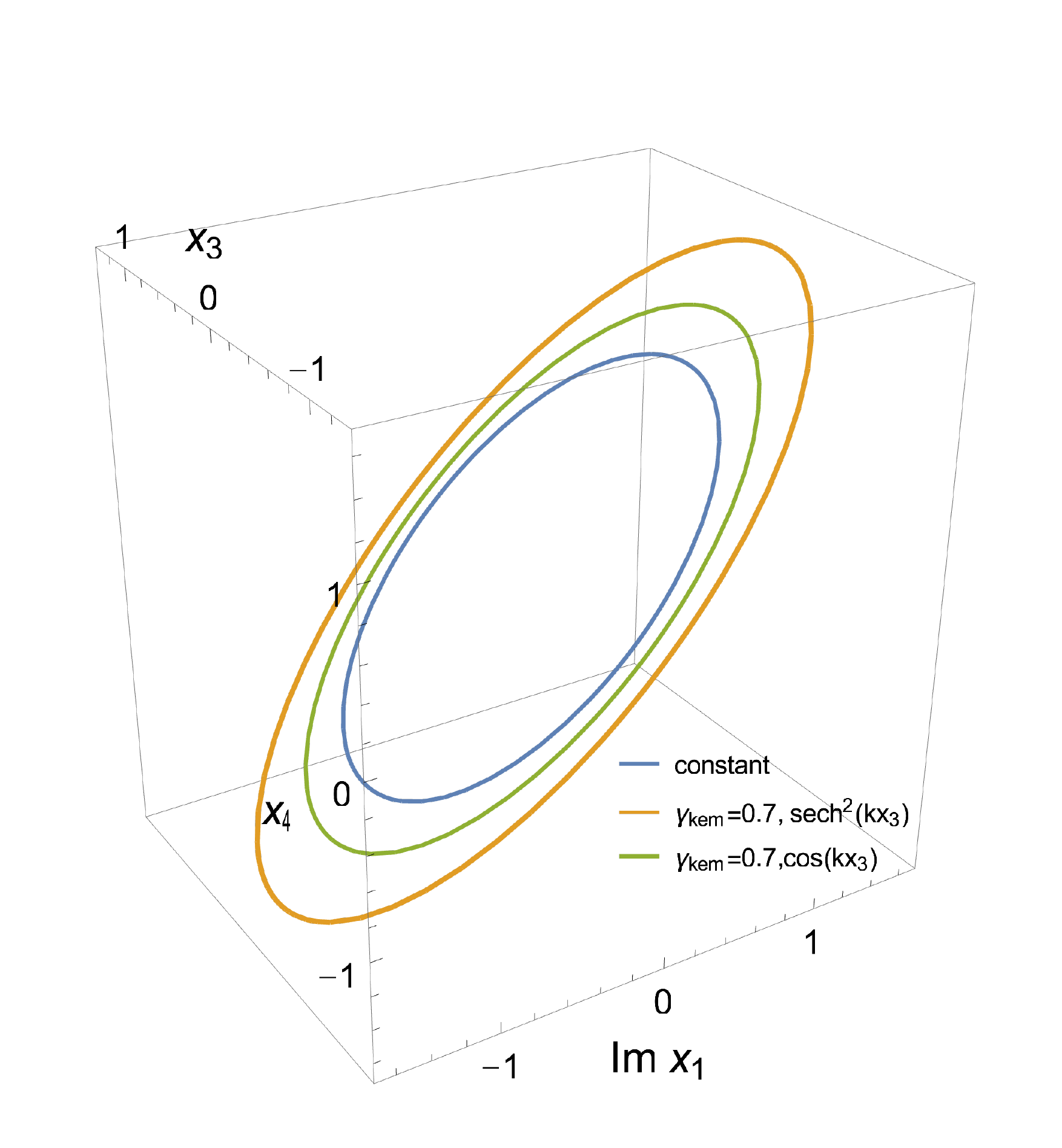}
\includegraphics[width=0.45\textwidth]{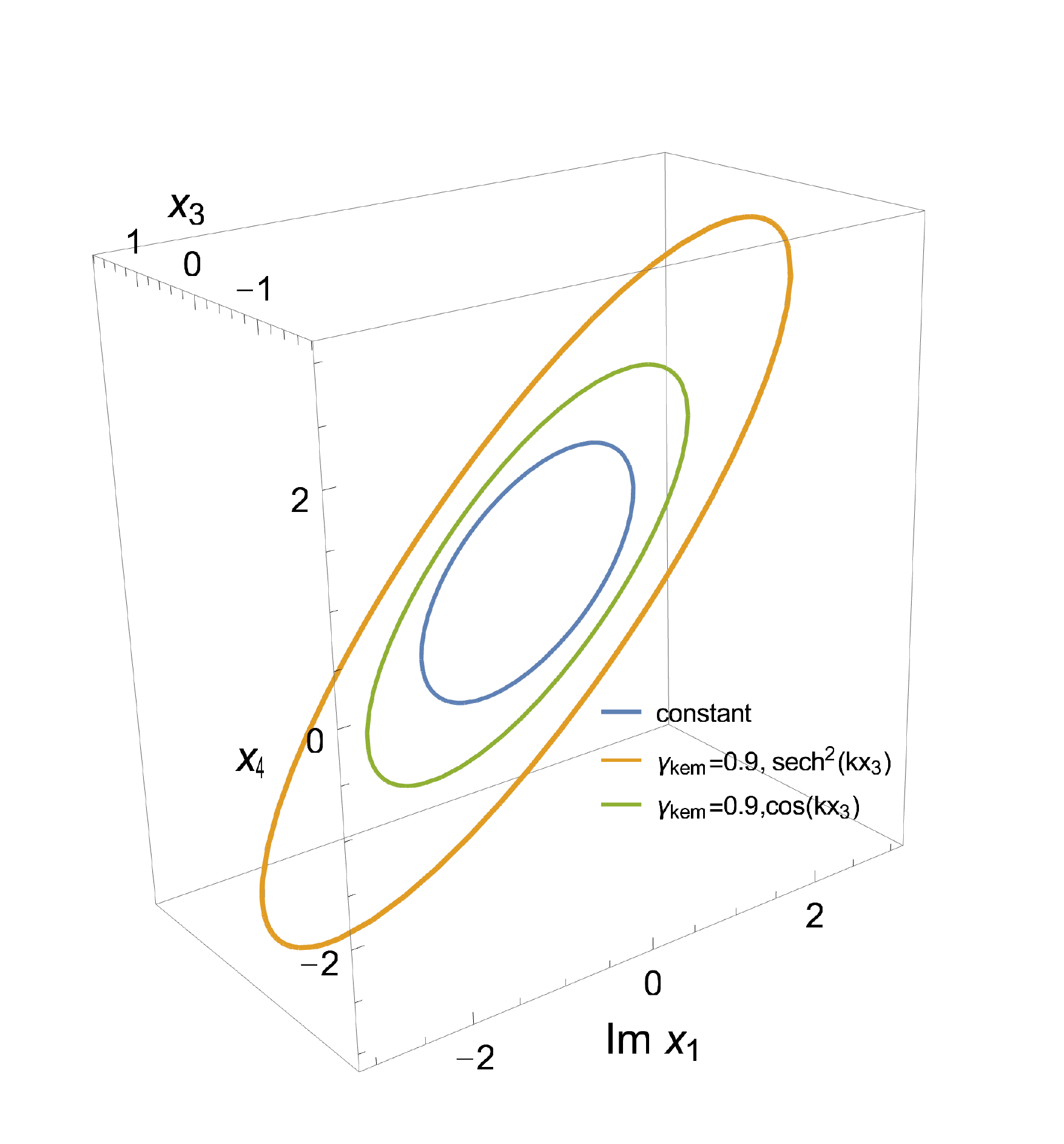}
\caption{Parametric plot of the stationary worldline instanton paths in the $({\rm Im}x_{1},~x_{3},~x_{4})$ space for the cases of the constant (blue), ${\rm sech}^{2}(kx_{3})$ (pink) and ${\rm cos}(kx_{3})$ (green) electric fields of strength $E$ and $B$. The paths are shown for various values of the parameter $\gamma_{keb}$(top right, top left, bottom right and bottom left for $0.1$, $0.5$, $0.7$ and $0.9$), ${\rm Im} x_{1}$, $x_{3}$ and $x_{4}$ have been expresses in units of $mB/e(E^{2}-B^{2})$, $m/e\sqrt{E^{2}-B^{2}}$ and $mE/e(E^{2}-B^{2})$.
}\label{fig:allpaths}
\end{figure*}
In this section, we discuss whether all of the instanton paths for electromagnetic fields as the Eqs.~\eqref{maxwellequatione}  and ~\eqref{maxwellequationb} are within the same plane in the $\left({\rm Im} x_{1},~x_{3},~x_{4} \right)$. To do so, we compare the instanton paths of different electromagnetic fields with different parameter $\gamma_{keb}$, as shown in Fig.~\ref{fig:allpaths}. We can find that the paths of the instantons are in the same plane for all  electromagnetic fields. As the parameter $\gamma_{keb} \rightarrow 0$, all paths tend to become the same elliptical shape, and the size of the instantons paths expands with $\gamma_{keb}$. For the same $\gamma_{keb}$, the size of instanton path that caused by constant electromagnetic fields is the smallest, ${\rm cos}(kx_{3})$ is in the middle, and ${\rm sech}^{2}(kx_{3})$ is the largest.

In the previous section, we discussed that for the same electromagnetic field, the paths of instantons are in the same plane. Therefore, all of the paths of instantons in our work are in the same plane indeed. For the same $\gamma_{keb}$, the instanton action $\mathcal{S}_{0}$ is the smallest for the constant electromagnetic field. However, the instanton actions are the middle and the largest for ${\rm cos}(kx_{3})$ and ${\rm sech}^{2}(kx_{3})$. According to Eq.~\eqref{probability}, for the same parameter $\gamma_{keb}$, the pair production probabilities  are the largest, middle and the smallest for constant, ${\rm cos}(kx_{3})$ and ${\rm sech}^{2}(kx_{3})$ electromagnetic fields. This prefactor can be deduced from the WKB approach or semiclassical method directly in Ref.~\cite{Dunne:2006st}, because the worldline instanton formalism can not provide prefactor. It is shown how the worldline instanton technique extends to spinor QED~\cite{Dunne:2005sx}. Therefore, we do not discuss this problem in this paper.

\section{Conclusion}\label{conclusion}

In conclusion, we have investigated the Schwinger effect in different space-dependent inhomogeneous electromagnetic fields with variation of parameter $\gamma_{keb}$ via the worldline instanton approach. It is found that all results would recover to that in Ref.~\cite{Dunne:2005sx} if $B=0$. Also we found that the range of loop paths of all instantons expand with $\gamma_{keb}$. All of the paths of instantons in our paper are in the same plane with various electromagnetic fields for different parameter $\gamma_{keb}$.

Moreover, the instanton action can be written generally as
\begin{align}\label{worldlinestandard}
\begin{split}
\mathcal{S}_{0}&= \frac{\pi mn}{k\gamma_{keb}} g(\gamma_{keb}),
\end{split}
\end{align}
where the function $g(\gamma_{keb})$ is defined as
\begin{align}\label{gg}
g(\gamma_{keb})&= \frac{2}{\pi} \int_{-1}^{1}dy \frac{\sqrt{1-y^{2}}}{|f^{'}|},
\end{align}
where $y=\frac{1}{\gamma_{keb}} f(\upsilon)$, $f^{'}(\upsilon)$ is to be reexpressed as a function of $y$. For the electromagnetic fields we discussed, the path of the instanton expands with $\gamma_{keb}$. At the same time, the instanton action and the tunneling time increase with $\gamma_{keb}$, the pair production rate decreases with $\gamma_{keb}$ decrease. Most importantly, the magnitudes of the instantons produced by various external electromagnetic fields are in the region between $\pi$ and $2\pi$ in normalized unit according to the analytical solutions of instantons action.

At last, for the pair-production time, we have identified two definitions and obtained the analytical expression of them. The distributions of the vacuum-decay times are similar to those shapes of the applied electromagnetic field modes for the same parameter $E/E_{cr}$. We find a nature of tunneling time for the worldline instanton approach that the tunneling time increases with the increase of parameter $\gamma_{keb}$ while $\gamma_{ke}$ and $k$ are fixed. Physically, the time it takes for an instanton to travel half loop path from $x^{-}_{3}$ to $x^{+}_{3}$ is the tunneling time. the larger the tunneling time, the smaller the pair creation. Interestingly, the tunneling time no longer depends on the $x_{3}$. Finally, we obtain the relationship between tunneling time and $x_{4}$
\begin{align}\label{tunnelingtimeafa}
	\mathcal{T}_{t}=2 x_{4}\left( 0 \right)= \left( x_{4}^{max} - x_{4}^{min}\right).
	\end{align}

These results suggest that the Schwinger effect in the space-dependent inhomogeneous electromagnetic fields is vital when we consider real laser pulses analytically. Our results may not only helping the analytical interpretation of pair production, but also has great prospects for the study on the essence of the pair-production time.
\begin{acknowledgments}
We would like to thank Mamutjan Ababekri for useful discussion and valuable comments. We are also grateful to Lie Juan Li and Li Wang for the critical reading of the manuscript and important discussions. This work was supported by the National Natural Science Foundation of China (NSFC) under Grant No. 11875007 and No. 11935008. The computation was carried out at the HSCC of the Beijing Normal University.
\end{acknowledgments}

\appendix

\section{Tunneling time in the QED}\label{app:Appa}

For the tunneling process, the particle energy is smaller than the potential energy of the particle so that the particle momentum is the imaginary number. It is convenient to calculate the tunneling time easily by using momentum of the instanton after perform Wick-rotation in the space-time $x_{\mu}=\left(t, -{\bm x}\right)$, where $x=|{\bm x}|=\sqrt{x_{1}^{2}+x_{2}^{2}+x_{3}^{2}}$, because the instanton momentum is the real value in the tunneling process. Thus, the relativistic momentum can be expressed with the instanton momentum via the Einstein's mass-energy relation
\begin{align}\label{instantonmomentum1}
p&=\sqrt{\mathcal{E}^{2}/c^{2}-m_{0}^{2}c^{2}}=i\sqrt{m_{0}^{2}c^{2}-\mathcal{E}^{2}/c^{2}}=iq,\\ \label{instantonmomentum2}
\mathcal{E}&=m c^{2}=m_{0} c^{2} / \sqrt{1 - \upsilon^{2}/c^{2} }=\gamma m_{0} c^{2},
\end{align}
where $p$, $\mathcal{E}$, $m_{0}$ and $m$ are the particle momentum, energy, rest mass and the relativistic mass of motion, $c$ is the speed of the light, $\upsilon$ denotes the magnitude of particle velocity ${\bm \upsilon}$ (${\bm \upsilon}=\left(\upsilon_{1}, \upsilon_{2}, \upsilon_{3}  \right)$), $\gamma$ is the relativistic factor. We define the instanton momentum $q$ in order to calculate the tunneling time, and the tunneling time can be written as~\cite{Landauer:1994zz}
\begin{align}\label{tunnelingtimeappendix3}
\mathcal{T}_{t}&=\int_{x_{-}}^{x_{+}} \frac{dx}{\upsilon},
\end{align}
where $\mathcal{T}_{t}$ and $x_{\pm}$ are the tunneling time and the classical turning points where $q\left(x_{\pm}\right)=0$~\cite{Greiner:2009}.

Now we calculate the tunneling time under constant electric field $E$ in order to compere with worldline instanton result conveniently. Thus, we can obtain the tunneling time in this case
\begin{align}\label{tunnelingtimeappendix4}
\mathcal{T}_{t}&=\int_{x_{-}}^{x_{+}} \frac{dx}{\frac{c}{\gamma} \sqrt{1- \left( \frac{W-eEx}{m_{0}c^{2}} \right)^{2}}},
\end{align}
here, $W$ denotes the energy of the instanton. If we define $u=\frac{W-eEx}{m_{0}c^{2}}$ in order to performing integral transformation, $\gamma=\frac{W-eEx}{m_{0}c^{2}} \equiv u$ because of the $\mathcal{E} \equiv \gamma m_{0}c^{2}= W - eEx$ under the constant electric field. Therefore, the tunneling time can be written as
\begin{align}\label{tunnelingtimeappendix5}
\begin{split}
\mathcal{T}_{t}&=2 \int_{0}^{1} \frac{ \left( {\Big |} -\frac{eE}{m_{0}c^{2}} {\Big |} \right)^{-1} u du}{c \sqrt{1- u^{2}}}=\frac{2m_{0}c}{eE} \int_{0}^{1} \frac{ u du}{\sqrt{1- u^{2}}}\\
&=\frac{2m_{0}c}{eE}.
\end{split}
\end{align}
We can find that the tunneling time $\mathcal{T}_{t}=2x_{4}\left(u=0\right)$ when $B=0$, $n=1$ and $\hbar=c=1$ in Eq.~\eqref{patheb3}. Therefore, it can be written as
\begin{align}\label{tunnelingtimeappendix6}
\mathcal{T}_{t}&\equiv \left( x_{4}^{max} - x_{4}^{min} \right).
\end{align}

\section{Tunneling time via the Hamilton-Jacobi equation}\label{app:Bppa}
We derive the same result Eq.~\eqref{tunnelingtimeappendix6} by using the Hamilton-Jacobi equation (HJE). From Eq. (16.11) in Ref.~\cite{Landau:1980}, the HJE can be written as
\begin{align}\label{tunnelingtimeappendixb1}
\left( \nabla \mathcal{S} -\frac{e}{c} {\bm A}  \right)^{2} -\frac{1}{c^{2}} \left( \frac{\partial \mathcal{S}}{\partial t} + e A_{4} \right)^{2} + m_{0}^{2}c^{2} =0,
\end{align}
where $A_{\mu}=(A_{4},~{\bm A})$, $\mathcal{S}$ is the action. The first and second terms represent particle energy and momentum according to the Einstein's mass-energy relation $p^{2}-\mathcal{E}^{2}/c^{2}+m_{0}^{2}c^{2}=0$. Thus, the tunneling time can be expressed by the instanton momentum $q$
\begin{align}\label{tunnelingtimeappendixb2}
\begin{split}
\mathcal{T}_{t}&=\int_{x_{-}}^{x_{+}} \frac{dx}{\upsilon}=2\int_{0}^{x_{+}} \frac{\mathcal{E}/c^{2} dx}{|p|}\\
&=2\int_{0}^{x_{+}} \frac{\mathcal{E}/c^{2} dx}{q}\\
&=2\int_{0}^{x_{+}} \frac{ \left( \frac{\partial \mathcal{S}}{\partial t} + e A_{4} \right) dx}{ c^{2} {\Big |} \left( \nabla \mathcal{S} -\frac{e}{c} {\bm A}  \right){\Big |}}.
\end{split}
\end{align}
After change the Einstein's mass-energy relation to $\mathcal{E}^{2}/m_{0}^{2}c^{4}-p^{2}/m_{0}^{2}c^{2}=\mathcal{E}^{2}/m_{0}^{2}c^{4}+q^{2}/m_{0}^{2}c^{2}=1$ form and two side multiply constant $a^{2}$ from Eq.~\eqref{constent}, we can obtain
\begin{align}\label{tunnelingtimeappendixb3}
\frac{\mathcal{E}^{2}a^{2}}{m_{0}^{2}c^{4}}+\frac{q^{2}a^{2}}{m_{0}^{2}c^{2}}=a^{2}\equiv \dot{x}^{2}_{1}+\dot{x}^{2}_{2}+\dot{x}^{2}_{3}+\dot{x}^{2}_{4}=\dot{x}^{2}+\dot{x}^{2}_{4}.
\end{align}
We can obtain the energy $\mathcal{E}$ and momentum $p$ by comparing with each terms in above equation
\begin{align}\label{tunnelingtimeappendixb4}
\mathcal{E}^{2}&=\left( \frac{\partial \mathcal{S}}{\partial t} + e A_{4} \right)^{2}=\frac{m_{0}^{2}c^{4}}{a^{2}} \dot{x}^{2}_{4},\\ \label{tunnelingtimeappendixb5}
p^{2}&=\left( \nabla \mathcal{S} -\frac{e}{c} {\bm A}  \right)^{2}=- \frac{m_{0}^{2}c^{2}}{a^{2}} \dot{x}^{2}\equiv -q^{2}.
\end{align}
After we take above equations into Eq.~\eqref{tunnelingtimeappendixb2}, the tunneling tome can be rewritten as
\begin{align}\label{tunnelingtimeappendixb6}
\begin{split}
\mathcal{T}_{t}&=2\int_{0}^{x_{+}} \frac{\dot{x}_{4}}{\dot{x}}dx=2\int_{1/4}^{0} \frac{\dot{x}_{4}}{\dot{x}}\frac{dx}{du}du=2\int_{1/4}^{0} \dot{x}_{4} du\\
&=2\left( x_{4}\left( 0 \right) -x_{4}\left( 1/4 \right) \right)=2 x_{4}\left( 0 \right)\equiv \left( x_{4}^{max} - x_{4}^{min} \right),
\end{split}
\end{align}
where $x_{4}^{min}$ and $x_{4}^{max}$ are maximum and minimum values for $x_{4}$. It is the same result Eq.~\eqref{tunnelingtimeappendix6}
, but Eq.~\eqref{tunnelingtimeappendixb6} does not depend on the external electromagnetic filed.

Most importantly, we can use this result for the space-time dependent inhomogeneous electromagnetic field
\begin{align}\label{tunnelingtimeappendixb7}
\mathcal{T}_{t}&=\int_{u\left(x_{-}\right)}^{u\left(x_{+}\right)} \dot{x}_{4} du.
\end{align}
The Eq.~\eqref{tunnelingtimeappendixb7} is general form for any field. We can prove Eq.~\eqref{tunnelingtimeappendixb7} via another convenient way, the tunneling time can be found by using instanton velocity
\begin{align}\label{tunnelingtimeappendixb8}
\mathcal{T}_{t}&=\int_{x_{-}}^{x_{+}} \frac{dx}{\upsilon}=\int_{x_{-}}^{x_{+}} \frac{dx}{\frac{dx}{dx_{4}}}=\int_{u\left(x_{-}\right)}^{u\left(x_{+}\right)} \dot{x}_{4} du.
\end{align}
This illustrates that the three results in the appendix are equivalent.

\appendix


\end{document}